\documentclass[pra,twocolumn,preprintnumbers,amsmath,amssymb,floatfix,superscriptaddress,showpacs]{revtex4}
\usepackage{graphicx}
\usepackage{graphics}
\usepackage{psfrag}
\usepackage{hyperref}
\usepackage{multirow}
\usepackage[sort&compress]{natbib}

\newcommand{\vare}{\varepsilon}
\newcommand{\rmi}{{\rm i}}

\begin{document}

\hypersetup{pdftitle={Quasi-long-range order in trapped two-dimensional Bose gases}}
\title{Quasi-long-range order in trapped two-dimensional Bose gases}

\author{Igor Boettcher}
\affiliation{Department of Physics, Simon Fraser University, Burnaby, British Columbia V5A 1S6, Canada}
\author{Markus Holzmann}
\affiliation{LPMMC, UMR 5493 of CNRS, Universit{\'e} Grenoble Alpes, BP 166, 38042 Grenoble, France}
\affiliation{Insitut Laue Langevin, BP 156, 38042 Grenoble, France}

\begin{abstract}
We study the fate of algebraic decay of correlations in a harmonically trapped two-dimensional degenerate Bose gas. The analysis is inspired by recent experiments on ultracold atoms where power-law correlations have been observed despite the presence of the external potential. We generalize the spin wave description of phase fluctuations to the trapped case and obtain an analytical expression for the one-body density matrix within this approximation. We show that algebraic decay of the central correlation function persists to lengths of about 20\% of the Thomas--Fermi radius. We establish that the trap-averaged correlation function decays algebraically with a strictly larger exponent weakly changing with trap size and find indications that the recently observed enhanced scaling exponents receive significant contributions from the normal component of the gas. We discuss radial and angular correlations and propose a local correlation approximation which captures the correlations very well. Our analysis goes beyond the usual local density approximation and the developed summation techniques constitute a powerful tool to investigate correlations in inhomogeneous systems.
\end{abstract}


\maketitle

With the achievement of quantum degeneracy in ultracold atom gases a new experimental platform for studying fundamental questions related to phase transitions and critical phenomena has appeared. In particular, correlation functions can be measured through interference and time-of-flight techniques \cite{Polkovnikov18042006,Hadzibabic2006,Clade2009,Tung2010,Plisson2011,PhysRevA.89.053612,Murthy2014}. A crucial aspect of the experiments, however, consists in the absence of translation invariance due to the presence of the external trapping potential. Consequently, correlations between points $\textbf{r}$ and $\textbf{r}'$ are not uniquely determined by the value of $\textbf{r}-\textbf{r}'$. This effect should be unimportant for short-range correlations, and, in fact, thermodynamic properties of ultracold gases are often very well-captured by a local density approximation. The situation changes drastically for a system at a critical point, where the correlation length diverges and thus competes with the inhomogeneity of the trapping potential.

Whereas the experimental preparation of critical systems typically requires a highly fine-tuned setup, they are rather effortlessly realized in two-dimensional (2D) systems whose low-energy excitations can be mapped onto an XY model. Examples apart from 2D ultracold quantum gases \cite{Dalibard2011,Desbuquois2012,PhysRevLett.114.230401,PhysRevLett.114.255302,PhysRevLett.115.010401} are given by thin Helium films \cite{Bishop1978}, layered magnets \cite{Durr1989,Ballentine1990}, and 2D exciton-polariton condensates \cite{Roumpos2012}. To quantify correlations in these systems we introduce the one-body density matrix $\rho(\textbf{r},\textbf{r}')=\langle \hat{\Phi}^\dagger(\textbf{r})\hat{\Phi}(\textbf{r}')\rangle$, where $\hat{\Phi}^\dagger(\textbf{r})$ is the creation operator for a particle at point $\textbf{r}$. For the spatially homogeneous case we then have $\rho_{\rm hom}(\textbf{r},\textbf{r}')=f(|\textbf{r}-\textbf{r}'|)$ with some function $f(r)$ due to translation and rotation invariance. The Mermin--Wagner--Hohenberg theorem \cite{Mermin1966,Hohenberg1967} forbids long-range order at any finite temperature such that $\lim_{r\to \infty}f(r)=0$. 
However, in the XY model an ordered low-temperature phase with infinite correlation length exists, where correlations decay with a temperature-dependent scaling exponent $\eta$ as $f(r) \sim r^{-\eta}$ for large $r$. 
Since $\eta$ is well below unity, this behavior of correlations is named 
quasi-long-range order (QLRO). Above a transition temperature, vortices disorder the system and QLRO is lost \cite{Holzmann2008,choi2013}. The described picture has been developed by Berezinskii and Kosterlitz, Thouless (BKT) \cite{Berezinskii1971,Berezinskii1972,Kosterlitz1973,Kosterlitz1974}.

The presence of the trap raises several conceptual questions towards the validity of the BKT picture in ultracold quantum gas experiments: On which length scales is QLRO visible? If samples bigger than the state of Texas are needed to reach the thermodynamic limit in the homogeneous system \cite{Holdsworth1994}, do we need ultracold atomic clouds covering the whole continent for the local density approximation to hold?
Does the formula $\eta = \frac{Mk_{\rm B}T}{2\pi \hbar^2 n_{\rm s}}$ for a Bose gas with mass $M$, temperature $T$, and superfluid density $n_{\rm s}$ imply a locally varying scaling exponent due to the inhomogeneous density? Can the 2D XY-model explain the large scaling exponent $\eta>1$ recently observed in experiment and Quantum Monte Carlo (QMC) computations \cite{PhysRevLett.115.010401}? Can we obtain the correlation function of the trapped system from the homogeneous one by a generalization of the local density approximation?

In this Rapid Communication, we address and answer all of these questions within the spin wave approximation for the phase fluctuations of the trapped gas. The main results are (1) the generalization of the textbook spin wave theory to a discrete collective mode spectrum, (2) the derivation of explicit analytical expressions for the one-body density matrix and first-order correlation functions that are suited for practical applications, and (3) the demonstration that the trap-averaged correlation function decays algebraically with an increased exponent. Our calculations provide a simple picture to the large exponents observed in experiment and simulation. We limit the derivation to the key steps and invite the mathematically interested reader to consult the detailed Supplemental Material (SM) \cite{SOM}.

\emph{Spin wave theory.} At sufficiently high temperatures, quantum corrections are small and the macroscopic properties of the trapped gas are described by the classical action
\begin{align}
 \label{eq1} S[\Phi] = \frac{1}{T} \int \mbox{d}^2r \Biggl[ \mbox{ }&\frac{|\nabla \Phi|^2}{2M}+\Bigl(-\mu+V(\textbf{r})\Bigr)|\Phi|^2+\frac{g}{2}|\Phi|^4\Biggr]
\end{align}
for the complex bosonic field $\Phi(\textbf{r})=A(\textbf{r})e^{\rmi \theta(\textbf{r})}$.  Herein, $\mu$ and $g$ are the chemical potential and coupling constant, and we use units such that $\hbar=k_{\rm B}=1$. We assume an isotropic and harmonic trapping potential $V(\textbf{r}) = \frac{M}{2}\omega^2r^2$. 
The action is stationary towards small changes in $A(\textbf{r})$ if $(-\frac{\nabla^2}{2M} -\mu+V+g A_0^2+\frac{(\nabla\theta)^2}{2M})A_0=0$ is satisfied.
We neglect gradient terms of the fields and obtain a Thomas--Fermi (TF) density profile given by $n(r) = A^2_0(\textbf{r}) =  n_0(1-r^2/R^2)$ with $n_0=\mu/g$ and TF radius $R=(2gn_0/M\omega^2)^{1/2}$. By neglecting deviations of $A(\textbf{r})$ from the stationary configuration we then arrive at the action for the phase $\theta(\textbf{r})$ given by 
\begin{align}
\label{eq2} S_{\rm ph}[\theta]= \frac{n_0}{2MT} \int_\lambda^{R} \mbox{d}^2r\ \Bigl(1-\frac{r^2}{R^2}\Bigr) (\nabla\theta)^2.
\end{align}
We explicate the length scales that delimit the validity of our description, given by the TF radius $R$ at large distances, and a microscopic scale $\lambda$, which is given by the thermal wavelength in cold atom experiments, or by the lattice spacing in spin models. For typical 2D ultracold quantum gases we have $R\simeq 100 \mu\text{m}$ and $\lambda\simeq 1 \mu\text{m}$, such that $R/\lambda\simeq100$ is large.

After partial integration the phase-only action can be written as quadratic form $S_{\rm ph}[\theta]=\frac{n_0}{2MT}(\theta,D_R\theta)$ with self-adjoint operator
\begin{align}
 \label{eq3}  D_R = -\Bigl(1-\frac{r^2}{R^2}\Bigr) \nabla^2 +\frac{2\textbf{r}\cdot\nabla}{R^2}
\end{align}
and scalar product $(f,g)=\int^R \mbox{d}^2r f(\textbf{r})g(\textbf{r})$. The normalized eigenfunctions of $D_R$ for $R<\infty$ are given by
\begin{align}
 \label{eq4} \theta_{nm}(\textbf{r}) = \sqrt{\frac{2n+|m|+1}{\pi}} s^{|m|/2} P_n^{(|m|,0)}(1-2s)e^{\rmi m \phi}
\end{align}
with $\textbf{r}=\binom{r\cos\phi}{r\sin \phi}$, $s=r^2/R^2$, $n\in\mathbb{N}_0$, $m\in \mathbb{Z}$, and Jacobi and Legendre polynomials $P_n^{(\alpha,\beta)}(x)$ and $P_n^{(0,0)}(x)=P_n(x)$, respectively \cite{AbrStegun,SOM}. The energy levels read
\begin{align}
 \label{eq5} \vare_{nm} = 2|m|+4n(n+|m|+1).
\end{align}
These energies match collective superfluid modes of harmonically trapped 2D Bose gases \cite{BoettchHydro}\footnote{Eq. (\ref{eq5}) is obtained from Eq. (6) in Ref. \cite{BoettchHydro} for $\alpha=1$, $d=2$, and $l=|m|$.}.
In fact, spin waves $\theta$ are related to the superfluid velocity by means of $\textbf{v}_{\rm s}=\frac{\hbar}{M}(\nabla\theta)$, i.e., they describe the same physical excitations \cite{PhysRevLett.77.1671,PhysRevLett.77.2360,2016arXiv160300434D}.

Since the phase-only action $S_{\rm ph}$ is quadratic, though with a nontrivial discrete spectrum, correlation functions are obtained via Gaussian integration. 
Within the spin wave approximation we have $ \rho(\textbf{r},\textbf{r}') = A_0(\textbf{r})A_0(\textbf{r}') \langle e^{\rmi[\theta(\textbf{r})-\theta(\textbf{r}')]}\rangle = A_0(\textbf{r})A_0(\textbf{r}') e^{-\frac{1}{2} \langle\Delta\theta(\textbf{r},\textbf{r}')^2\rangle}$ with $\Delta\theta(\textbf{r},\textbf{r}')=\theta(\textbf{r})-\theta(\textbf{r}')$ and
\begin{align}
 \label{eq6} \langle\Delta\theta(\textbf{r},\textbf{r}')^2\rangle =  \frac{MT}{n_0}\sum_{n,m}\ \hspace{-2mm}^\prime\ \frac{|\theta_{nm}(\textbf{r})-\theta_{nm}(\textbf{r}')|^2}{\vare_{nm}}.
\end{align}
In the homogeneous limit ($R=\infty$) we have $D_\infty=-\nabla^2$ and the eigenfunctions are plane waves $\theta_{\textbf{q}}(\textbf{r})=e^{\rmi \textbf{q}\cdot\textbf{r}}$ with energies $\vare_{\textbf{q}}=q^2$, and
\begin{align}
 \nonumber \langle\Delta\theta(\textbf{r},\textbf{r}')^2\rangle_{\rm hom} &=  \frac{MT}{n_0}\int^{\lambda^{-1}} \frac{\mbox{d}^2q}{(2\pi)^2}  \frac{|\theta_{\textbf{q}}(\textbf{r})-\theta_{\textbf{q}}(\textbf{r}')|^2}{\vare_{\textbf{q}}}\\
  \label{eq7} &\simeq\eta_0\log\Bigl(\frac{|\textbf{r}-\textbf{r}'|^2}{\lambda^2}\Bigr)\ \text{for}\ |\textbf{r}-\textbf{r}'|\gg \lambda,
\end{align}
with $\eta_0 = \frac{MT}{2\pi n_0}$ \cite{HerbutBook,AltlandBook}. Due to $s^{m/2}P_n^{(m,0)}(1-2s) \sim J_m(2n \sqrt{s})$ and $\vare_{nm}\sim 4n^2$ for large $n$ and small $s$, we recover this formula for $R\to \infty$. The homogeneous case is recalled in the SM \cite{SOM}.  
In the homogeneous system, spin wave theory breaks down for $\eta_0\simeq 0.25$ \cite{Kosterlitz1973,Kosterlitz1974}. In a trap it is applicable to the superfluid core in the center of the atomic cloud. We estimate its radius $r$ from $n(r)\lambda^2 > 4$ as $r^2/R^2< 1-4\eta_0$. For $\eta_0=0.10$ ($0.20$) we have $r< 0.8R$ ($0.4R$), which is sufficiently large to observe the effects studied here. Much of the elegance of BKT theory  stems from replacing the integral in Eq. (\ref{eq7}) by the logarithm, which is valid for all relevant separations, but not exact.

\emph{Evaluation of the sum.} Just as the momentum integration in the homogeneous case is limited by $q\lambda\leq 1$, also Eq. (\ref{eq6}) has an ultraviolet cutoff on possible values of $(n,m)$, indicated by the prime. In fact, besides $(n,m)\neq(0,0)$, the values of $n$ and $m$ are limited from above according to $\vare_{nm}\leq\vare_{\rm max} \sim (R/\lambda)^2$. 
In the following we choose summation cutoffs $N= e^{-\gamma}(R/\lambda),\ M = e^{-\gamma}(R/\lambda)^2$
with Euler's constant $\gamma\simeq0.577$ \cite{SOM}. We can then write
\begin{align}
 \nonumber \langle \Delta\theta(\textbf{r},\textbf{r}')^2\rangle&= \frac{\eta_0}{2}\Bigl[ F_0^{(N)}(s,s) - 2 F_0^{(N)}(s,s') +F_0^{(N)}(s',s')\Bigr]\\
 \nonumber  &\hspace{-3mm}+\eta_0 \sum_{m=1}^M \Bigl[F_m^{(N_m)}(s,s)-2 \cos (m \Delta \phi) F_m^{(N_m)}(s,s')\\
 \label{eq9} &\hspace{-3mm} +F_m^{(N_m)}(s',s')\Bigr]
\end{align}
with $N_{m}=-\frac{m+1}{2}+\frac{\sqrt{2M+m^2+1}}{2}$ and
\begin{align}
 \nonumber F_0^{(N)}(s,s') &= \sum_{n=1}^N \frac{2n+1}{n(n+1)} P_n(1-2s)P_n(1-2s'),\\
\nonumber F_m^{(N_m)}(s,s') &= \sum_{n=0}^{N_m} \frac{2n+m+1}{\frac{m}{2}+n(n+m+1)} (ss')^{m/2} \\
 \label{eq11} &\times P_n^{(m,0)}(1-2s)P_n^{(m,0)}(1-2s').
\end{align}
 Here and in the following we denote
\begin{align}
 \nonumber &\textbf{r} = \begin{pmatrix}r\cos\phi\\r\sin\phi\end{pmatrix},\ s=\frac{r^2}{R^2},\ \Delta\phi= \phi-\phi',\\
 \label{eq12}  &s_>=\text{max}(s,s'),\ s_<=\text{min}(s,s'),\ \hat{\lambda}=\frac{\lambda}{R}.
\end{align}
Equation (\ref{eq9}) can be implemented numerically and thus allows the computation of the phase fluctuations in a trap for arbitrary values of $R/\lambda$. Note in particular that $\langle\Delta\theta(\textbf{r},\textbf{r})^2\rangle =0$. If $R/\lambda$ is large, however, analytical approximation formulas can be derived from the $N,M\to \infty$ limits.

Evaluating Eq. (\ref{eq11}) for $N,N_m=\infty$ relies on the observation that for a self-adjoint operator $\hat{O}$ with eigenvalues $\vare_{i}$ and eigenfunctions $f_i(s)$, the function $g(s,s')=-\sum_i\frac{f_i(s)f_i(s')}{\vare_i}$ is the Green function of the operator due to the completeness of the $f_i$. Accordingly, if we know the Green function by some other means, we also know the result of the summation \cite{EnglisPeetre,Gustavsson}. Now, for fixed $m$ the operator $D_R$ is of Sturm--Liouville-type and thus its Green function is given by $u(s_<)v(s_>)$, where $u,v$ are zero modes of the operator. With this approach we find
\begin{align}
 \label{eq13} F_0^{(\infty)}(s,s') &= -1-\log[s_{>}(1-s_{<})],\\
 \label{eq14} F_m^{(\infty)}(s,s') &= \frac{1}{m} \Bigl(\frac{s_{<}}{s_{>}}\Bigr)^{m/2} u_m(s_{<})v_m(s_{>})
\end{align}
for $0<s,s'<1$. We have $u_m(s) = {}_2F_1(a_m,b_m,m+1,s)$ and $v_m(s) = \frac{\Gamma(a_m)\Gamma(b_m)}{\Gamma(m)} {}_2F_1(1-a_m,1-b_m,1,1-s)$
with hypergeometric function $_2F_1(a,b,c,z)$  and $a_m =\frac{m+1}{2}+\frac{\sqrt{m^2+1}}{2}$, $b_m =\frac{m+1}{2}-\frac{\sqrt{m^2+1}}{2}$ such that $a_m+b_m=m+1$ and $a_mb_m=m/2$ \cite{AbrStegun,SOM}. The key properties of the functions $u_m$ and $v_m$ are $u_m(0)=v_m(0)=1$ and
\begin{align}
 \label{eq17} \lim_{m\to \infty}u_m(s) = \lim_{m\to \infty}v_m(s) = \frac{1}{\sqrt{1-s}}.
\end{align}
We define the remainder function
\begin{align}
 \label{eq18} \mathcal{R}(s,s',\Delta\phi) &= \sum_{m=1}^{M_0} \frac{1}{m}\Bigl(\frac{s_<}{s_>}\Bigr)^{m/2}\cos(m\Delta\phi)\\
 \nonumber  &\times \Bigl(u_m(s_<)v_m(s_>)-\frac{1}{\sqrt{(1-s)(1-s')}}\Bigr),
\end{align}
which converges quickly in $M_0$ and is well-approximated by summing up the first $M_0= 10$ terms. The careful limit $N,M\to \infty$ in Eq. (\ref{eq9}) is presented in the SM \cite{SOM}.

\emph{Correlation functions.} The phase fluctuations due to Eq. (\ref{eq9}) for $R/\lambda\gg10$ are given by
\begin{widetext}
\begin{align}
 \nonumber \langle\Delta\theta(\textbf{r},\textbf{r}')^2\rangle &\simeq \frac{\eta_0}{2}\Bigl(1+\frac{1}{1-s_>}\Bigr)\log\Bigl(\frac{\bar{s}_>}{\hat{\lambda}^2(1-s_>)}\Bigr)-\frac{\eta_0}{2}\Bigl(1-\frac{1}{1-s_<}\Bigr)\log\Bigl(\frac{\bar{s}_<}{\hat{\lambda}^2(1-s_<)}\Bigr)\\
 \label{eq19}&+\frac{\eta_0}{\sqrt{(1-s)(1-s')}}\log\Bigl(\frac{\Delta \bar{\textbf{r}}^2}{\bar{s}_>}\Bigr)+\eta_0\Bigl(\mathcal{R}(s,s,0)+\mathcal{R}(s',s',0)-2\mathcal{R}(s,s',\Delta\phi)\Bigr)
\end{align}
\end{widetext}
with $\bar{s} = \max\{s,\hat{\lambda}^2\},\ \Delta \bar{\textbf{r}}^2 = \max\{\frac{|\textbf{r}-\textbf{r}'|^2}{R^2}, \hat{\lambda}^2(1-s)\}$.
The latter two definitions give a proper way to treat the singularities that occur when $r,r',|\textbf{r}-\textbf{r}'|\to 0$. Their origin is already visible in the homogeneous limit (\ref{eq7}): Whereas the integral is well-defined for all $\textbf{r},\textbf{r}'$, the logarithm expression requires a meaningful procedure of how to treat the ultraviolet singularities. The remainder term involving the $\mathcal{R}$-functions in Eq. (\ref{eq19}) can be neglected for most cases of interest. The homogeneous limit $\eta_0\log(|\textbf{r}-\textbf{r}'|^2/\lambda^2)$ is recovered for $\hat{\lambda}^2\ll s,s'\ll1$, i.e., $\lambda\ll r,r'\ll R$.

\begin{figure}[t!]
\centering
\begin{minipage}{0.46\textwidth}
\includegraphics[width=8.2cm]{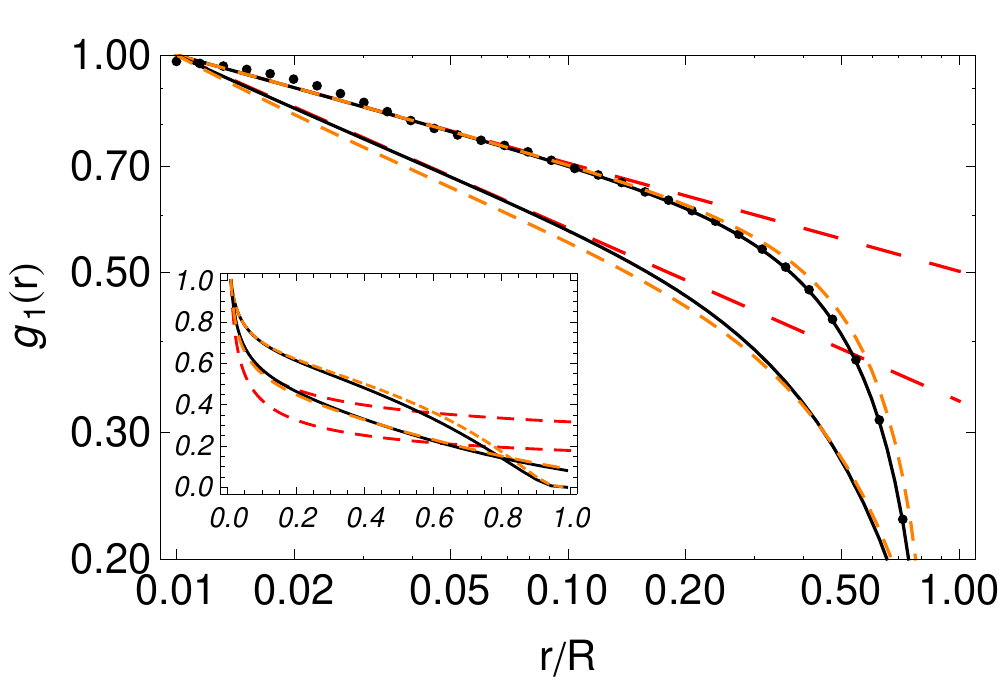}
\includegraphics[width=8.2cm]{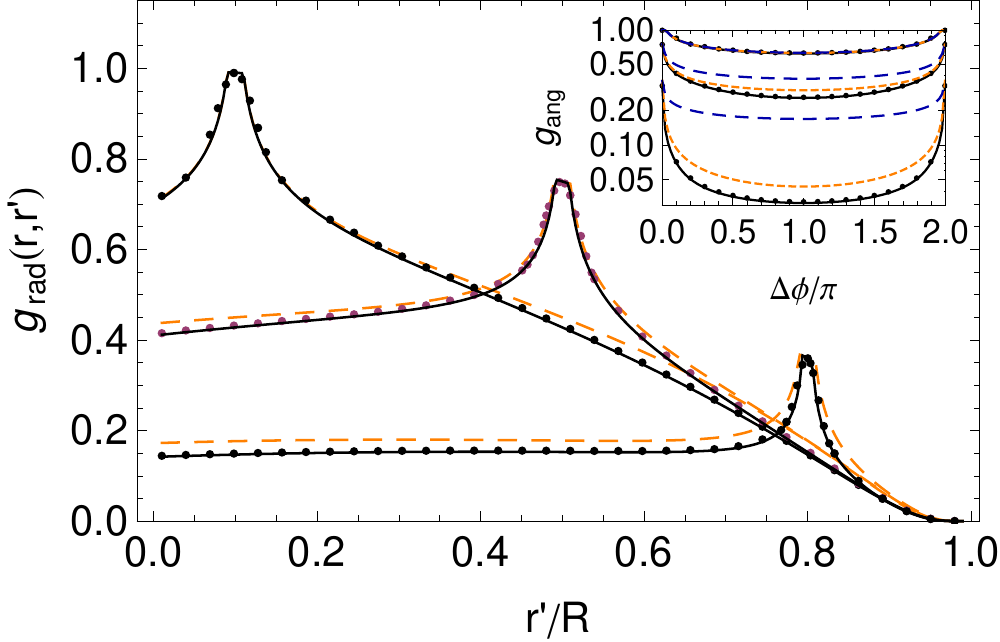}
\caption{Selection of correlation functions for $R/\lambda=100$ and $\eta_0=0.15$ in terms of $\rho(r,r',\Delta\phi)=\langle\Phi^*(\textbf{r})\Phi(\textbf{r}')\rangle$.
Black data represents the sum from Eq. (\ref{eq6}) (points) and the analytic result from Eq. (\ref{eq19}) (solid line). The LCA from Eq. (\ref{eq26}) is indicated by the dashed orange lines. \emph{Upper panel.} Central correlation function $g_1(r)=\rho(r,0,0)$ (upper curve) decaying algebraically with exponent $\eta_0$, and trap-averaged correlation function $\bar{g}_1(r)$ (lower curve) decaying with strictly larger exponent $\eta_{\rm trap}>\eta_0$. Long-dashed red lines correspond to $(r/\lambda)^{-\eta}$ with $\eta=\eta_0$ and $\eta=1.6\eta_0$, respectively. The inset shows the same data on a linear scale. \emph{Lower panel.} Radial correlation function $g_{\rm rad}(r,r')=\rho(r,r',0)$ for fixed $r/R=0.1,0.5,0.8$ (from left to right). Note that the peaks heights are $(1-r^2/R^2)$. The inset shows the angular correlation function $g_{\rm ang}(r,\Delta\phi)=\rho(r,r,\Delta\phi)$ for the same values of $r/R$ (from top to bottom). The long-dashed blue lines show $(1-r^2/R^2)(|\textbf{r}-\textbf{r}'|/\lambda)^{-\eta_0}$, which deviates considerably.
}
\label{Fig1}
\end{minipage}
\end{figure}

The central correlation function $g_1(r)=\rho(\textbf{r},0)$ derived from Eq. (\ref{eq19}) is given by
\begin{align}
 \label{eq22} g_1(r) = n_0(1-s)^{\frac{1}{2}[1+\eta(s)]}\Bigl(\frac{r}{\lambda}\Bigr)^{-\eta(s)} e^{-\frac{\eta_0}{2}\mathcal{R}(s,s,0)}
\end{align}
with $\eta(s)=\frac{\eta_0}{2}\bigl(1+\frac{1}{1-s}\bigr)$.
The factor $e^{-\frac{\eta_0}{2}\mathcal{R}}$ is close to unity and may be neglected. The algebraic decay with $\eta_0$ for small $s\ll 1$ agrees with the central correlation functions obtained in Refs. \cite{Petrov2000B,PhysRevA.83.035602,PhysRevB.88.024517}. The angular correlations for $r=r'$ read
\begin{align}
 \nonumber \langle\Delta\theta(r,r,\Delta\phi)^2\rangle &= \frac{\eta_0}{1-s}\log\Bigl(\frac{2s(1-\cos\Delta\phi)}{\hat{\lambda}^2(1-s)}\Bigr)\\
 \label{eq23} &+2\eta_0[\mathcal{R}(s,s,0)-\mathcal{R}(s,s,\Delta\phi)]
\end{align}
for $\Delta\phi^2\geq \hat{\lambda}^2(1-s)/s$. (This corresponds to $|\textbf{r}-\textbf{r}'|^2\geq\lambda^2(1-s)$. For $\Delta\phi\to0$ the correlations vanish.) In Fig. \ref{Fig1} we plot a selection of correlation functions.

\begin{figure}[t!]
\centering
\includegraphics[width=7.5cm]{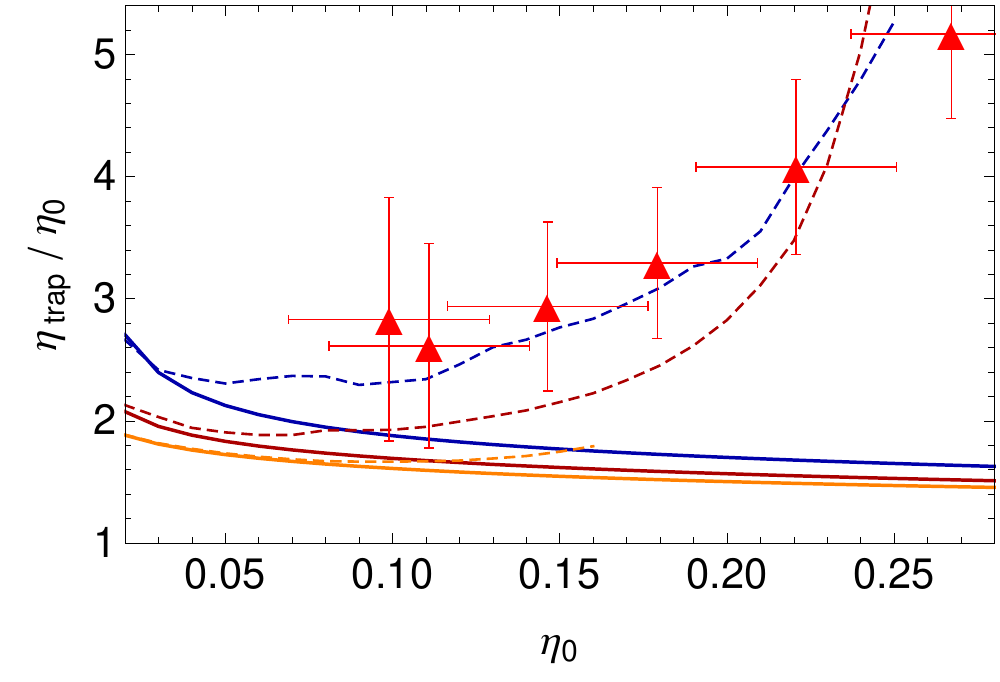}
\caption{Scaling exponent $\eta_{\rm trap}$ extracted from $\bar{g}_1(r)\sim (r/\lambda)^{-\eta_{\rm trap}}$ for $R/\lambda=20,\ 100,\ 1000$ (blue, dark red, orange). The solid lines give the result of applying Eq. (\ref{eq19}) to the whole cloud. As a first estimate of the influence of the superfluid transition we replace the phase correlations in the outer regions of the cloud by an exponential decay with correlation length $\lambda$. This yields an increase of $\eta_{\rm trap}$ for small $R/\lambda$ (dashed lines). The red triangles correspond to the QMC data from Ref. \cite{PhysRevLett.115.010401} for the quasi-2D gas with $\tilde{g}=0.60$ and $R/\lambda\approx 50$, with $\eta_0$ determined from the scaling of $g_1(r)$. The error bars estimate the fitting error of $\eta_0$, which is substantial due to its small value.}
\label{Fig2}
\end{figure}

\emph{Applications.} To obtain a useful estimate of the trapped correlation function from the homogeneous result, we write Eq. (\ref{eq7}) as
\begin{align}
 \label{eq25} \langle\Delta\theta(\textbf{r},\textbf{r}')^2\rangle_{\rm hom} = 2G(\textbf{r},\textbf{r}')-G(\textbf{r},\textbf{r})-G(\textbf{r}',\textbf{r}')
\end{align}
with $G(\textbf{r},\textbf{r}')=\frac{\eta_0}{2}\log(|\textbf{r}-\textbf{r}'|^2/R^2),\ G(\textbf{r},\textbf{r})=\frac{\eta_0}{2}\log(\hat{\lambda}^2)$, the Green function of $-\frac{MT}{n_0}\nabla^2$. In each $G$ separately we apply a local density approximation $\eta_0 \to\eta(\textbf{r},\textbf{r}')=\frac{MT}{2\pi \sqrt{n(r)n(r')}}=\frac{\eta_0}{\sqrt{(1-s)(1-s')}}$ and obtain
\begin{align}
 \nonumber \langle\Delta\theta(\textbf{r},\textbf{r}')^2\rangle_{\rm LCA} &=  \frac{\eta_0}{2}\Bigl[\frac{1}{1-s}+\frac{1}{1-s'}\Bigr]\log\Bigl(\frac{1}{\hat{\lambda}^2}\Bigr)\\
 \label{eq26} &+\frac{\eta_0}{\sqrt{(1-s)(1-s')}} \log\Bigl(\frac{|\textbf{r}-\textbf{r}'|^2}{R^2}\Bigr).
\end{align}
We name this particular procedure local correlation approximation (LCA), and observe from Fig. \ref{Fig1} that it works sufficiently well to approximate the actual result. In contrast, the approximation $\rho(\textbf{r},\textbf{r}')\approx A(\textbf{r})A(\textbf{r}')(|\textbf{r}-\textbf{r}'|/\lambda)^{-\eta_0}$ fails considerably.

The Fourier transform of the momentum distribution defines the trap-averaged correlation function \cite{Cohen-Tannoudji2011,PhysRevLett.115.010401}
\begin{align}
 \label{eq24} \bar{g}_1(r) = \int \mbox{d}^2x\ \rho(\textbf{x},\textbf{r}+\textbf{x}).
\end{align}
In a translation invariant setting we have $\bar{g}_1(r)\propto g_1(r)$, whereas both functions differ in the trapped case. 
As shown in Figs. \ref{Fig1} and \ref{Fig2}, $\bar{g}_1(r)$ decays algebraically for $\lambda \leq r \ll R$ with an increased exponent $\eta_{\rm trap}\simeq (1.5$-$2)\eta_0$, which weakly depends on $R/\lambda$. In the SM \cite{SOM} we analytically estimate $\eta_{\rm trap}$, which supports the exceptionally weak $R/\lambda$-dependence. Although $\eta_{\rm trap} \to \eta_0$ for infinitely large systems, it would require absurd values of $R/\lambda$.

The spin wave description only applies to the superfluid core of the cloud. Hence, applying Eq. (\ref{eq19}) to the whole cloud becomes less accurate for increasing values of $\eta_0$ and underestimates $\eta_{\rm trap}$. To estimate the effect of the normal gas on the trap-averaging we set $\langle e^{\rmi(\theta(\textbf{r})-\theta(\textbf{r}'))}\rangle= e^{-|\textbf{r}-\textbf{r}'|/\lambda}$ for $(r,r')\notin [0,r_{\rm s}] \times [0,r_{\rm s}]$ with approximate core radius $r_{\rm s}=(1-4\eta_0)^{1/2}R$ \cite{SOM}. Assuming the correlation length in the normal gas to coincide with $\lambda$ should give the right order of magnitude of the effect. Computing $\bar{g}_1(r)$ with this modification of $\rho(\textbf{r},\textbf{r}')$, we find an increased scaling exponent $\eta_{\rm trap}$ which is slightly above the spin wave prediction for $\eta_0 \lesssim 0.15$. Increasing $\eta_0$ further, we obtain exponents up to $\eta_{\rm trap}\approx 1.5$ for small $R/\lambda\lesssim 100$, whereas the algebraic decay disappears for larger $R/\lambda$ in the typical fitting range $r\lesssim 0.1R$. Since most experiments are in the first regime, an enhanced scaling exponent is likely to be measured. Neglecting the spin wave part and only keeping exponential correlations, no algebraic decay of $\bar{g}_1(r)$ is observed.

In order to relate to the enhanced exponents found in Ref. \cite{PhysRevLett.115.010401}, we compare to the QMC results where $\eta_0$ is determined by the decay of the central correlation function \cite{Holzmann2010}\footnote{Experimental results of Ref. \cite{PhysRevLett.115.010401} contain an additional large imaging contribution at low temperatures.}. As shown in Fig \ref{Fig2}, we find good qualitative agreement with the QMC results when including the exponential decay of correlations in the normal component. Hence we identify the superfluid core, a significant normal component, and a ratio of $R/\lambda \sim 10^2$ as a possible explanation for the large universal transition exponent $\eta_{\rm trap, c} \simeq 1.4$. 
Although it is possible to include more quantitative knowledge of the quasi-2D equation of state \cite{2008EL.....8230001H,Holzmann2010,PhysRevA.66.043608,PhysRevLett.116.045303,PhysRevLett.115.240401} into the LCA approximation, it is clear that our qualitative findings are robust against these refinements. A full quantitative analysis of the BKT transition in the trapped gas is left for future work.

In conclusion, we have generalized the spin wave theory of phase fluctuations towards the trapped 2D Bose gas, analytically computed correlation functions, and applied this to obtain the trap-averaged correlation function. Our analysis provides a natural explanation for the unexpected large exponents observed in Ref. \cite{PhysRevLett.115.010401}. It should be possible to apply the method to other dimensions and polytropic equations of state, just as it is the case for superfluid hydrodynamic modes \cite{PhysRevA.68.043610,PhysRevLett.93.040402,BoettchHydro}. We proposed an LCA prescription which can be further developed by considering these other mode spectra. An improved understanding of correlations in inhomogeneous systems is also mandatory for conducting experiments on universality in disordered systems \cite{PhysRevA.85.033602,beeler2012,PhysRevLett.111.050406} and situations far from equilibrium \cite{2011arXiv1112.1204M,PhysRevX.5.011017,PhysRevX.5.041028}.

\emph{Acknowledgments.} We gratefully acknowledge inspiring discussions with A. B\"{o}ttcher, R. Ganesh, I. F. Herbut, P. A. Murthy, and M. M. Scherer. IB acknowledges funding by the European Research Council (ERC) under Advanced Grant No. 290623 and the German Research Foundation (DFG) under Grant No. BO 4640/1-1. MH was supported by ANR-12-BS04-0022-01, ANR-13-JS01-0005-01 and DIM Nano'K from R\'{e}gion \^{I}le-de-France.

\bibliographystyle{apsrev4-1}
\bibliography{refs_phase} 

\cleardoublepage

\setcounter{equation}{0}
\renewcommand{\theequation}{S\arabic{equation}}

\begin{center}
\textbf{\Large Supplemental Material}
\end{center}

\tableofcontents

\section{Computation of the correlation function}\label{AppComp}
In this section we derive the formula for the phase fluctuations in detail. The presentation provides all the intermediate steps that have been left out in the main text for reasons of brevity.

We start by solving the eigenvalue problem for the operator 
\begin{align}
 D_R = -\Bigl(1-\frac{r^2}{R^2}\Bigr) \nabla^2 +\frac{2\textbf{r}\cdot\nabla}{R^2}
\end{align}
appearing in the phase-only action
\begin{align}
 \label{swAction} S_{\rm ph}[\theta] = \frac{n_0}{2MT}\int_\lambda^R \mbox{d}^2r\ \Bigl(1-\frac{r^2}{R^2}\Bigr)(\nabla\theta)^2 = \frac{n_0}{2MT}(\theta,D_R\theta).
\end{align}
We introduce dimensionless variables $\hat{r}=r/R$ and $\hat{D}=R^2D_R$ to obtain the eigenvalue problem
\begin{align}
 \label{sw9} \hat{D} \theta_\vare = \vare \theta_\vare.
\end{align}
We employ a polar coordinate Ansatz $\theta_\vare(\textbf{r})=\sum_{m=-\infty}^\infty \theta_{\vare m}(\hat{r})e^{\rmi m \phi}$ to arrive at
\begin{align}
  \label{sw10}  & -(1-\hat{r}^2) \Bigl(\frac{\mbox{d}^2}{\mbox{d} \hat{r}^2}+\frac{1}{\hat{r}}\frac{\mbox{d}}{\mbox{d} \hat{r}}-\frac{m^2}{\hat{r}^2}\Bigr)\theta_{\vare m} +2\hat{r}\frac{\mbox{d}}{\mbox{d} \hat{r}}  \theta_{\vare m}=\vare \theta_{\vare m}.
\end{align}
Note that since $\hat{D}$ is invariant under $\textbf{r}\to -\textbf{r}$, the eigenvalues only depends on $|m|$. With the new coordinate
\begin{align}
 \label{sw11} s = \hat{r}^2 =  r^2/R^2
\end{align}
we eventually arrive at $\hat{D}^{(m)}\theta_{\vare m}=\vare \theta_{\vare m}$ with
\begin{align}
 \label{sw12} \hat{D}^{(m)} = -\frac{\mbox{d}}{\mbox{d}s}\Bigl[4s(1-s)\frac{\mbox{d}}{\mbox{d}s}\Bigr] +m^2 \frac{1-s}{s}.
\end{align}
Note that this operator is of Sturm--Liouville form, see Eq. (\ref{sturm1}). The eigenfunctions of $\hat{D}^{(m)}$ are easily seen to be of the form $s^{|m|/2}P_n^{(|m|,0)}(1-2s)$ with $P_n^{(\alpha,\beta)}$ being the Jacobi polynomials, see Eq. (\ref{jac7}). The corresponding eigenvalues are given by
\begin{align}
 \label{sw13} \vare_{nm} = 2|m|+4n(n+|m|+1)
\end{align}
with $n=0,1,2,\dots$ such that $(n,m)\neq(0,0)$. For $m=0$ the solution reduces to the Legendre polynomial $P_n(1-2s)$.  In App. \ref{AppLegendre} we collect relevant properties of the Legendre and Jacobi polynomials.

We define the normalized eigenfunctions
\begin{align}
 \label{sw14} \theta_{nm}(\textbf{r}) = \sqrt{\frac{2n+|m|+1}{\pi}} s^{|m|/2} P_n^{(|m|,0)|}(1-2s)e^{\rmi m \phi}
\end{align}
such that
\begin{align}
 \label{sw15} \hat{D}\theta_{nm} = R^2D_R\theta_{nm}=\vare_{nm}\theta_{nm},
\end{align}
and, due to Eq. (\ref{jac3}), $(\theta_{nm},\theta_{n'm'})=R^2 \delta_{nn'}\delta_{m,-m'}$. Due to the completeness of the Jacobi polynomials, every \emph{regular} phase configuration $\theta(\textbf{r})$ can be expanded according to
\begin{align}
 \label{sw17} \theta(\textbf{r}) = \sum_{n,m} a_{nm} \theta_{nm}(\textbf{r}).
\end{align}
Since $\theta(\textbf{r})$ is real we have $a_{nm}^*=a_{n,-m}$. We arrive at the regular phase-only action
\begin{align}
 \label{sw18} S_{\rm ph}[\theta] = \frac{n_0}{2MT} \sum_{n,m}\ \hspace{-2mm}^\prime\ \vare_{nm}|a_{nm}|^2.
\end{align}
The smallest and largest length scales $\lambda$ and $R$ introduced in Eq. (\ref{swAction}) manifest in lower and upper boundaries on the summation (indicated by a prime). The implications of this are addressed below in detail.

We now consider the one-body density matrix
\begin{align}
 \label{cor1} \rho(\textbf{r},\textbf{r}') = \langle \Phi^*(\textbf{r})\Phi(\textbf{r}')\rangle.
\end{align}
Note that since $\langle \Phi(\textbf{r}) \rangle=0$ this correlation function coincides with the connected correlation function $\langle \Phi^*(\textbf{r})\Phi(\textbf{r}')\rangle-\langle \Phi^*(\textbf{r})\rangle\langle\Phi(\textbf{r}')\rangle$. We write $\Phi(\textbf{r})=A(\textbf{r})e^{\rmi \theta(\textbf{r})}$ and assume $\theta(\textbf{r})$ to be regular. Approximating $A(\textbf{r})$ by the stationary solution $A_0(\textbf{r})=\sqrt{n(r)}=\sqrt{n_0(1-s)}$ we arrive at
\begin{align}
  \label{cor2} \rho(\textbf{r},\textbf{r}') &= \sqrt{n(r)n(r')}\langle e^{-\rmi(\theta(\textbf{r})-\theta(\textbf{r}'))}\rangle
\end{align}
with
\begin{align}
 \label{cor3} \langle \mathcal{O}\rangle = \frac{\int\mbox{D}\theta\ \mathcal{O} \ e^{-S_{\rm ph}[\theta]}}{\int\mbox{D}\theta\ e^{-S_{\rm ph}[\theta]}}.
\end{align}
Since $S_{\rm ph}[\theta]$ is quadratic and $\langle\theta(\textbf{r})\rangle=0$ we have
\begin{align}
 \label{cor4} \langle e^{-\rmi(\theta(\textbf{r})-\theta(\textbf{r}'))}\rangle = e^{-\frac{1}{2}\langle[\theta(\textbf{r})-\theta(\textbf{r}')]^2\rangle}
\end{align}
according to the rules of Gaussian integration. 

We proceed by computing 
\begin{align}
\langle \Delta\theta(\textbf{r},\textbf{r}')^2\rangle = \langle[\theta(\textbf{r})-\theta(\textbf{r}')]^2\rangle
\end{align}
with the Gaussian action $S_{\rm ph}$. We write the functional measure as $\mbox{D}\theta = \prod_{\nu,\mu}\int\mbox{d}a_{\nu\mu}$ and find
\begin{align}
\nonumber  &\langle[\theta(\textbf{r})-\theta(\textbf{r}')]^2\rangle \\
 \nonumber &\hspace{4mm}= \frac{\int \mbox{D}\theta\ (\theta(\textbf{r})-\theta(\textbf{r}'))(\theta(\textbf{r})-\theta(\textbf{r}'))e^{-S_{\rm ph}[\theta]}}{\int \mbox{D}\theta\ e^{-S_{\rm ph}[\theta]}}\\
 \nonumber  &\hspace{4mm}=\sum_{n,m}\sum_{n',m'} (\theta_{nm}(\textbf{r})-\theta_{nm}(\textbf{r}')) (\theta_{n'm'}(\textbf{r})-\theta_{n'm'}(\textbf{r}'))\\ 
 \label{comp1}  &\hspace{4mm}\ \times \prod_{\nu,\mu} \frac{\int\mbox{d}a_{\nu\mu} \ a_{nm}a_{n'm'}e^{-\frac{1}{2}\gamma_{\nu\mu} a_{\nu\mu}a_{\nu,-\mu}}}{\int\mbox{d}a_{\nu\mu} \ e^{-\frac{1}{2}\gamma_{\nu\mu} a_{\nu\mu}a_{\nu,-\mu}}}
\end{align}
with $\gamma_{nm}=n_0\vare_{nm}/MT$. In the product, all terms except one yield unity. The nontrivial one is $\delta_{nn'}\delta_{m,-m'}\frac{1}{\gamma_{nm}}$ due to
\begin{align}
 \label{comp2} \frac{\int \mbox{d}a \ a^2 e^{-\frac{1}{2}a\gamma a}}{\int \mbox{d}a \ e^{-\frac{1}{2}a\gamma a}}=\frac{1}{\gamma}
\end{align}
for a real variable $a$. This eventually yields
\begin{align}
 \label{cor8} \langle \Delta\theta(\textbf{r},\textbf{r}')^2\rangle  =  \frac{MT}{n_0}\sum_{n,m}\ \hspace{-2mm}^\prime\ \frac{|\theta_{nm}(\textbf{r})-\theta_{nm}(\textbf{r}')|^2}{\vare_{nm}}.
\end{align}

The macroscopic description of the trapped Bose gas in terms of the phase-only action is restricted to length scales $r\gg \lambda$. This implies an ultraviolet energy cutoff according to $E_{\rm max} \sim \lambda^{-2}$. The energies $\vare_{nm}=2|m|+4n(n+|m|+1)$ are thus  restricted according to
\begin{align}
 \label{cr7} 0 < \vare_{nm} < \vare_{\rm max} \sim \Bigl(\frac{R}{\lambda}\Bigr)^2
\end{align}
with $\vare_{\rm max}=R^2E_{\rm max}$. For $m=0$ the largest term in the $n$-sum is 
\begin{align}
 \label{cr8} N=N_0 = \frac{-1+\sqrt{1+\vare_{\rm max}}}{2}  \simeq \frac{\sqrt{\vare_{\rm max}}}{2} \sim \Bigl(\frac{R}{\lambda}\Bigr).
\end{align}
The largest value of $m$, denoted by $M$, is attained for $n=0$ and is found to be
\begin{align}
 \label{cr9} M= \frac{\vare_{\rm max}}{2} \sim \Bigl(\frac{R}{\lambda}\Bigr)^2.
\end{align}
If $0<m<M$, the possible values of $n$ are limited from above by $N_m<N_0$. Given $m>0$ we have
\begin{align}
 2m+ 4N_m(N_m+m+1) < \vare_{\rm max}
\end{align}
for the largest value of $n=N_m$. Thus
\begin{align}
 \nonumber N_m &= -\frac{m+1}{2}+\frac{\sqrt{\vare_{\rm max}+m^2+1}}{2}\\
 \label{cr10} &= -\frac{m+1}{2}+\frac{\sqrt{2M+m^2+1}}{2}.
\end{align}
In the following we choose
\begin{align}
 \label{cr11} N = e^{-\gamma} \Bigl(\frac{R}{\lambda}\Bigr),\ M = e^{-\gamma} \Bigl(\frac{R}{\lambda}\Bigr)^2
\end{align}
with Euler's constant $\gamma\simeq0.577$. For large $N$ and $M$ this implies
\begin{align}
 \label{cr12} 2H_N = \log\Bigl(\frac{1}{\hat{\lambda}^2}\Bigr),\ H_M = \log\Bigl(\frac{1}{\hat{\lambda}^2}\Bigr)
\end{align}
with harmonic number $H_n=\sum_{\nu=1}^n\frac{1}{\nu}$. We will see below that fixing $N$ and $M$ according to Eqs. (\ref{cr12}) leads to convenient expressions for the phase fluctuations by trading $N,M$ for $\hat{\lambda}$. Note, however, that there is no $\vare_{\rm max}$ such that Eqs. (\ref{cr11}) can be true at the same time. On the other hand, due to $e^{-\gamma}\simeq 0.56 \simeq \frac{1}{2}$, Eqs. (\ref{cr11}) are close approximations to $\vare_{\rm max}=(R/\lambda)^2$, i.e., with a proportionality factor of unity in $E_{\rm max}\sim \lambda^{-2}$.

We now write the summation formula like in the main text as
\begin{align}
 \nonumber \langle \Delta\theta(\textbf{r},\textbf{r}')^2\rangle&= \frac{\eta_0}{2}\Bigl[ F_0^{(N)}(s,s) - 2 F_0^{(N)}(s,s') +F_0^{(N)}(s',s')\Bigr]\\
 \nonumber  &\hspace{-3mm}+\eta_0 \sum_{m=1}^M \Bigl[F_m^{(N_m)}(s,s)-2 \cos (m \Delta \phi) F_m^{(N_m)}(s,s')\\
 \label{cr13} &\hspace{-3mm} +F_m^{(N_m)}(s',s')\Bigr]
\end{align}
with
\begin{align}
 \nonumber F_0^{(N)}(s,s') &= \sum_{n=1}^N \frac{2n+1}{n(n+1)} P_n(1-2s)P_n(1-2s'),\\
\nonumber F_m^{(N_m)}(s,s') &= \sum_{n=0}^{N_m} \frac{2n+m+1}{\frac{m}{2}+n(n+m+1)} (ss')^{m/2} \\
 \label{cr14} &\times P_n^{(m,0)}(1-2s)P_n^{(m,0)}(1-2s').
\end{align}
We have
\begin{align}
 \label{cr15} F_0^{(\infty)}(s,s') &= -1-\log[s_{>}(1-s_{<})], \\
 \label{cr16} F_{m>0}^{(\infty)}(s,s') &= \frac{1}{m} \Bigl(\frac{s_{<}}{s_{>}}\Bigr)^{m/2} u_m(s_{<})v_m(s_{>})
\end{align}
with the functions $u_m(s)$ and $v_m(s)$ as in Eqs. (\ref{sum21}) and (\ref{sum22}), see App. \ref{AppSums}. We cannot, however, simply replace the finite sums with the infinite series as, depending on the arguments $(s, s', \Delta\phi)$, the sums can be ill-defined.

\begin{figure}[t]
\centering
\begin{minipage}{0.46\textwidth}
\includegraphics[width=7cm]{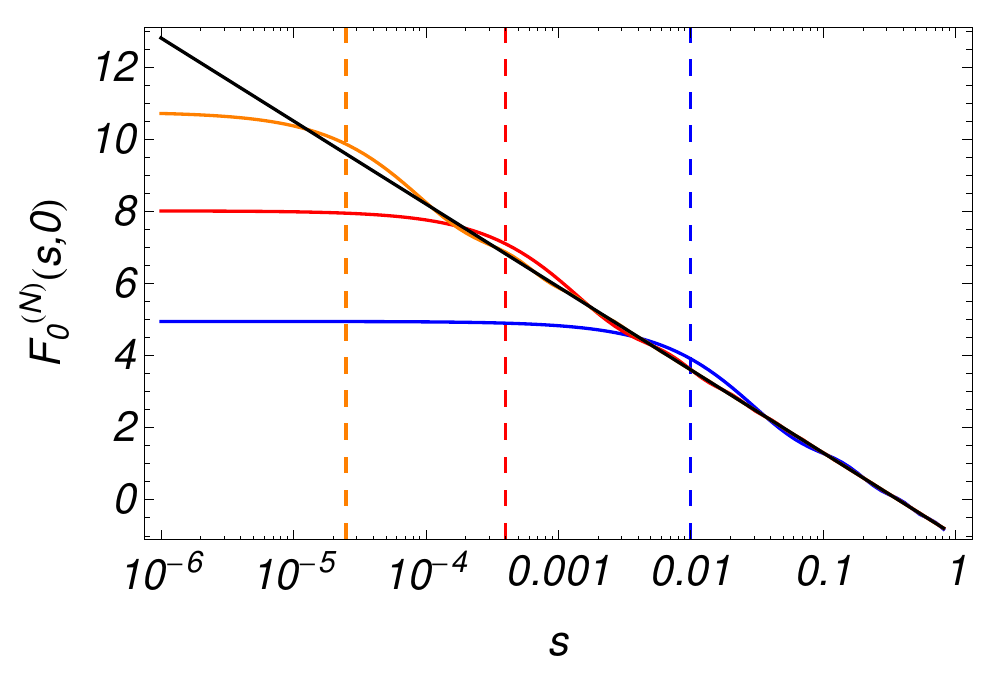}
\includegraphics[width=7cm]{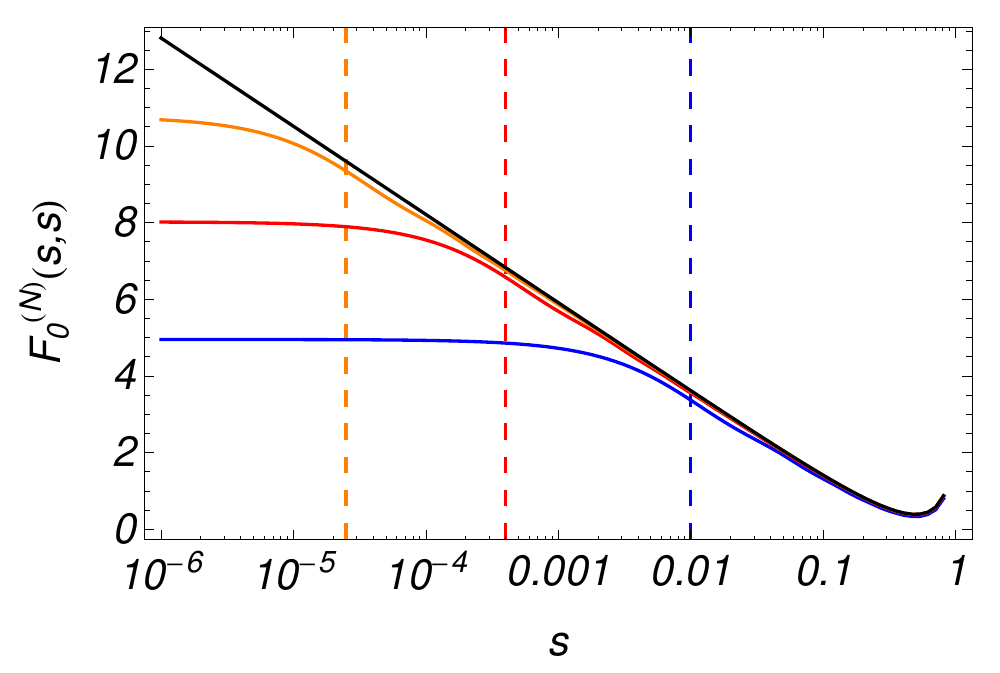}
\caption{From top to bottom: The functions $F_0^{(\infty)}(s,s')$ (black curve) and $F_0^{(N)}(s,s')$ for $N=200$ (orange curve), $N=50$ (red curve), and $N=10$ (blue curve). In the upper panel we show the case of $s'=0$, in the lower one $s'=s$. The vertical grid lines indicate $1/N^2$ for the corresponding coloured curve. We observe that the $N=\infty$ result is matched for $s\gtrsim 1/N^2$, which implies $r\gtrsim \lambda$. Hence, for all practical purposes we can approximate $F_0^{(N)}$ by $F_0^{(\infty)}$.}
\label{FigCross}
\end{minipage}
\end{figure}

We first consider the case of $m=0$. The functions $F_0^{(N)}(s,0)$ and $F_0^{(N)}(s,s)$ for both $N<\infty$ and $N=\infty$ are displayed in Fig. \ref{FigCross}. We observe that $F^{(\infty)}_0$ describes the sum for $s\gtrsim 1/N^2$. Using $N\simeq R/\lambda$ this amounts to $r\gtrsim \lambda$. Consequently, within the inherent limitations of the spin wave description of the Bose gas, we can always safely replace $F_0^{(N)}(s,s')\to F_0^{(\infty)}(s,s')$ in Eq. (\ref{cr13}) as long as $s,s'\neq 0$. For $s=s'=0$, on the other hand, we have
\begin{align}
 \label{cr17} F_0^{(N)}(0,0) = \sum_{n=1}^N \frac{2n+1}{n(n+1)} = 2H_N -1 +\frac{1}{N+1}.
\end{align}
For large $N$, choosing $N=e^{-\gamma}(R/\lambda)$ as in Eq. (\ref{cr11}), we then have
\begin{align}
 \label{cr18} F_0^{(N)}(0,0) \simeq  -1-\log \hat{\lambda}^2.
\end{align}
On the other hand, this coincides with $F_0^{(\infty)}(s,0)$ for $s=\hat{\lambda}^2$, i.e., $r=\lambda$. We see that there are two points of view on the regularization of the correlation functions: Either we fix $N<\infty$ and keep the summation formula. Then all values of $s$ are allowed. Or we replace the sum by the infinite series result (i.e. the logarithm), but are then limited to values $s\geq \hat{\lambda}^2$. Both approaches can be matched for practical purposes by replacing
\begin{align}
 \label{cr19} s \to \bar{s} = \max\{ s,\hat{\lambda}^2\}.
\end{align}
We then have
\begin{align}
 \label{cr20} F_0^{(N)}(s,s') &\simeq -1-\log[\bar{s}_{>}(1-s_{<})]
\end{align}
for all values of $s,s'$. Thereby we neglect finite corrections of order $\mathcal{O}(\hat{\lambda}^2)$ and only keep diverging contributions.

For $m>0$ the situation is more involved. We first consider the non-pathological cases of $|\textbf{r}- \textbf{r}'|\gg \lambda $ and $r,r'\gg \lambda$. We then find
\begin{align}
 \nonumber &\sum_{m=1}^M \cos(m\Delta\phi) F_m^{(N_m)}(s,s') \simeq  \sum_{m=1}^M \cos(m\Delta\phi) F_m^{(\infty)}(s,s')\\
 \nonumber &=\frac{1}{\sqrt{(1-s)(1-s')}}\sum_{m=1}^M \frac{1}{m} \Bigl(\frac{s_{<}}{s_{>}}\Bigr)^{m/2} \cos(m\Delta\phi)\\
 \nonumber  &+\sum_{m=1}^M \frac{1}{m} \Bigl(\frac{s_{<}}{s_{>}}\Bigr)^{m/2}\cos(m\Delta\phi) \\
 \label{cr23} &\times\Bigl(u_m(s_{<})v_m(s_{>})-\frac{1}{\sqrt{(1-s)(1-s')}}\Bigr)
\end{align}
for very large $M=\mathcal{O}(1000)$. Such large values are natural due to the power of two appearing in $M \sim (R/\lambda)^2$. For instance, $R/\lambda=100$ together with Eq. (\ref{cr11}) yields $M=5600$. In Eq. (\ref{cr23}) we added and subtracted the limits of $u_m$ and $v_m$ for $m\to \infty$. In the first sum we send $M\to \infty$ and make use of
\begin{align}
 \nonumber &\sum_{m=1}^\infty \frac{1}{m} \cos(m\Delta \phi) \Bigl(\frac{s_{<}}{s_{>}}\Bigr)^{m/2} \\
 \nonumber & = \frac{1}{2} \sum_{m=1}^\infty \frac{1}{m}\Bigl((\sqrt{x} e^{\rmi \Delta\phi})^m+(\sqrt{x}e^{-\rmi \Delta \phi})^m\Bigr)\\
 \nonumber &= \frac{1}{2} \Bigl[-\log(1-\sqrt{x}e^{\rmi \Delta\phi})-\log(1-\sqrt{x}e^{-\rmi \Delta\phi})\Bigr]\\
 \nonumber &= -\frac{1}{2}  \log\Bigl(1-2\sqrt{x}\cos \Delta\phi +x\Bigr)\\
  \label{cr24} &=-\frac{1}{2} \log\Bigl( \frac{|\textbf{r}-\textbf{r}'|^2}{r_{>}^2}\Bigr)
\end{align}
with $x=s_</s_>$. We identify the remaining terms in Eq. (\ref{cr23}) as
\begin{align}
 \label{cr25} \mathcal{R}(s,s',\Delta\phi) &= \sum_{m=1}^{M_0} \frac{1}{m}\Bigl(\frac{s_<}{s_>}\Bigr)^{m/2}\cos(m\Delta\phi)\\
 \nonumber  &\times \Bigl(u_m(s_<)v_m(s_>)-\frac{1}{\sqrt{(1-s)(1-s')}}\Bigr),
\end{align}
where $M_0$ is a small number of order $\mathcal{O}(10)$ as the sum converges fast. For practical purposes $M_0=10$ is a sufficient choice, but also $M_0=5$ works in most cases. Often the function $\mathcal{R}$ can be neglected altogether in comparison to the other terms in the correlation function. Note the enormous reduction in computational effort: We reduced the sum with $M\times N \simeq 5000 \times 50$ terms to an analytical contribution and a finite sum involving $M_0\simeq 10$ terms. We summarize the findings of this paragraph as
\begin{align}
 \nonumber &\sum_{m=1}^M \cos(m\Delta\phi) F_m^{(N_m)}(s,s') \\
 \label{cr26} &\simeq -\frac{1}{2\sqrt{(1-s)(1-s')}} \log\Bigl( \frac{|\textbf{r}-\textbf{r}'|^2}{r_{>}^2}\Bigr)+\mathcal{R}(s,s',\Delta\phi)
\end{align}
for $|\textbf{r}-\textbf{r}'|\gg\lambda$.

In the case that one of the arguments of the sum vanishes, say $s'\to 0$, we have
\begin{align}
 \label{cr27} \sum_{m=1}^M \cos(m\Delta\phi) F_m^{(N_m)}(s,0)=0
\end{align}
for every finite $M$. Since $\log(|\textbf{r}-\textbf{r}'|^2/r_>^2)=\log(r^2/r^2)=0$ and $\mathcal{R}(s,0,0)=0$ for $s'\to 0$, we conclude that Eq. (\ref{cr26}) is still valid.

In the case that $\textbf{r}=\textbf{r}'$ it is easily verified numerically that $\sum_{m=1}^M F_m^{(N_m)}(s,s)\neq \sum_{m=1}^M F_m^{(\infty)}(s,s)$ even for very large $M$. However, the result of the sum can be obtained as follows: First note that the term $\sum_{m=1}^M F_m^{(N_m)}(s,s)$ \emph{always} appears in the formula for $\langle\Delta(\textbf{r},\textbf{r}')^2\rangle$ in Eq. (\ref{cr13}). For small $r,r'\ll R$, however, the latter must reproduce the translational invariant homogeneous result $\eta_0\log(|\textbf{r}-\textbf{r}'|^2/\lambda^2)$. Choosing $\hat{\lambda}^2\ll s,s'\ll 1$ but keeping $|\textbf{r}-\textbf{r}'|\gg \lambda$ we apply the results of Eqs. (\ref{cr20}) and (\ref{cr26}) to write
\begin{align}
 \nonumber \langle\Delta\theta(\textbf{r},\textbf{r}')^2\rangle \stackrel{s,s'\ll1}{\longrightarrow} &\frac{\eta_0}{2}\log\Bigl(\frac{r_>^2}{r_<^2}\Bigr)+\eta_0\log\Bigl(\frac{|\textbf{r}-\textbf{r}'|^2}{r_>^2}\Bigr)\\
 \label{cr28}&+\eta_0\sum_{m=1}^M\Bigl(F_m^{(N_m)}(s,s)+F_m^{(N_m)}(s',s')\Bigr).
\end{align}
Consequently,
\begin{align}
 \label{cr29} \sum_{m=1}^MF_m^{(N_m)}(s,s)\stackrel{s\ll1}{\longrightarrow} \frac{1}{2} \log\Bigl(\frac{s}{\hat{\lambda}^2}\Bigr) = \frac{1}{2}\log s +\frac{1}{2}H_M.
\end{align}
We observe that, indeed, the logarithmic singularity for $\hat{\lambda}^2\ll s \ll 1$ is not captured by $\sum_{m=1}^M F_m^{(\infty)}(s,s)= \frac{H_M}{1-s} + \mathcal{R}(s,s,0)$. However, we can make an elaborate guess how Eq. (\ref{cr29}) generalizes to arbitrary $s$. We verify numerically that 
\begin{align}
 \label{cr30} \sum_{m=1}^MF_m^{(N_m)}(s,s) &\simeq \frac{1}{2(1-s)}\log\Bigl(\frac{s}{\hat{\lambda}^2(1-s)}\Bigr)+\mathcal{R}(s,s,0)
\end{align}
is an excellent approximation. Of course, it would be more satisfying to have an analytical understanding of the logarithmic singularities appearing in the sum, but we leave that to future work. We attribute them to the presence of the cutoff $N_m$ instead of $N\sim (R/\lambda)$, so that replacing $F_m^{(N_m)}\to F_m^{(\infty)}$ misses some terms that eventually diverge logarithmically when summed over $m$. Note that Eq. (\ref{cr30}) can be obtained from Eq. (\ref{cr26}) by replacing $|\textbf{r}-\textbf{r}'|^2\to \lambda^2(1-s)$.

The last case that needs to be studied is the angular correlation for $r=r'$, i.e.,
\begin{align}
 \label{cr31} \sum_{m=1}^M \cos(m\Delta\phi) F_m^{(N_m)}(s,s).
\end{align}
For $\Delta\phi\to 0$ we need to recover Eq. (\ref{cr30}). On the other hand, if $\Delta\phi>0$ is sufficiently large, Eq. (\ref{cr26}) should be valid, giving $-\frac{1}{2(1-s)}\log[2(1-\cos\Delta\phi)]+\mathcal{R}(s,s,\Delta\phi)$. Both results are matched for a minimal angle $\Delta\phi_\lambda$ which is given by
\begin{align}
 \label{cr32} \Delta\phi_\lambda^2 = \frac{\hat{\lambda}^2(1-s)}{s}.
\end{align}
However, this again corresponds to $|\textbf{r}-\textbf{r}'|^2=\lambda^2(1-s)$. We conclude that \emph{all} possible choices for $\textbf{r}$ and $\textbf{r}'$ are captured by the formula
\begin{align}
 \nonumber &\sum_{m=1}^M \cos(m\Delta\phi) F_m^{(N_m)}(s,s') \\
 \label{cr33} &\simeq -\frac{1}{2\sqrt{(1-s)(1-s')}} \log\Bigl( \frac{\Delta \bar{\textbf{r}}^2}{s_>}\Bigr)+\mathcal{R}(s,s',\Delta\phi)
\end{align}
with
\begin{align}
 \label{cr34} \Delta \bar{\textbf{r}}^2 = \max\Bigl\{ \frac{|\textbf{r}-\textbf{r}'|^2}{R^2},\hat{\lambda}^2(1-s)\Bigr\}.
\end{align}
We arrive at
\begin{align}
 \nonumber \langle\Delta\theta(\textbf{r},\textbf{r}')^2\rangle &= \frac{\eta_0}{2}\log\Bigl(\frac{\bar{s}_>(1-s_<)}{\bar{s}_<(1-s_>)}\Bigr)\\
 \nonumber &+\frac{\eta_0}{2(1-s)}\log\Bigl(\frac{\bar{s}}{\hat{\lambda}^2(1-s)}\Bigr)\\
  \label{cr35}  &+\frac{\eta_0}{2(1-s')}\log\Bigl(\frac{\bar{s}'}{\hat{\lambda}^2(1-s')}\Bigr)\\
 \nonumber &+\frac{\eta_0}{\sqrt{(1-s)(1-s')}}\log\Bigl(\frac{\Delta\bar{\textbf{r}}^2}{\bar{s}_>}\Bigr)\\
 \nonumber &+\eta_0\Bigl(\mathcal{R}(s,s,0)+\mathcal{R}(s',s',0)-2\mathcal{R}(s,s',\Delta\phi)\Bigr).
\end{align}
This result can be rewritten to yield the formula displayed in the main text. 

Note that it should be possible to relate the formula for $\langle\Delta\theta(\textbf{r},\textbf{r}')^2\rangle$ to the Green function of the operator $D_R$. However, we did not succeed in this direction and postpone it to future work.

\section{Homogeneous limit}\label{SecHom}

Several findings of the present analysis can be understood in close analogy to the homogeneous situation without a trapping potential, which is also covered in many textbooks on critical phenomena \cite{HerbutBook,AltlandBook}. In this section we give a brief reminder on the homogeneous limit formulas, first in terms of the  energy labels $q_x,q_y$, but also in the radially symmetric variables $q,m$, which is conceptually closer to our computation in the trap. 

As already pointed out in the main text we have $D_{R\to\infty}=-\nabla^2$. The eigenfunctions of the operator $D_\infty$ are given by $\theta_{\textbf{q}}(\textbf{r})=e^{\rmi\textbf{q}\cdot\textbf{r}}$ with eigenenergies $\vare_{\textbf{q}}=\textbf{q}^2=q_x^2+q_y^2$. We have the orthogonality and completeness relations
\begin{align}
 \label{hom1} \int\mbox{d}^2r\ \theta_{\textbf{q}}^*(\textbf{r}) \theta_{\textbf{q}'}(\textbf{r}) &= (2\pi)^2\delta^{(2)}(\textbf{q}-\textbf{q}'),\\
 \label{hom2} \int\mbox{d}^2q\ \theta_{\textbf{q}}^*(\textbf{r})\theta_{\textbf{q}}(\textbf{r}') &= (2\pi)^2\delta^{(2)}(\textbf{r}-\textbf{r}').
\end{align}
Consequently we can write $\theta(\textbf{r})=\int \frac{d^2q}{(2\pi)^2}\ a_{\textbf{q}}\theta_{\textbf{q}}(\textbf{r})$ with $a^*_{\textbf{q}}=a_{-\textbf{q}}$ for every regular configuration and have
\begin{align}
 \label{hom3} S_{\rm ph}[\theta] = \frac{n_0}{2MT} \int^\Lambda \frac{\mbox{d}^2q}{(2\pi)^2}\ \textbf{q}^2|a_{\textbf{q}}|^2.
\end{align}
We made explicit that the momentum integration has to be equipped with a proper ultraviolet cutoff $\Lambda\sim \lambda^{-1}$. From a calculation analogous to Eq. (\ref{comp1}) we arrive at
\begin{align}
 \label{hom4} \langle\Delta\theta(\textbf{r},\textbf{r}')^2\rangle_{\rm hom} = \frac{MT}{n_0} \int^\Lambda \frac{\mbox{d}^2q}{(2\pi)^2} \frac{|e^{\rmi\textbf{q}\cdot\textbf{r}}-e^{\rmi \textbf{q}\cdot\textbf{r}'}|^2}{\textbf{q}^2}.
\end{align}

\begin{figure}[t]
\centering
\includegraphics[width=8cm]{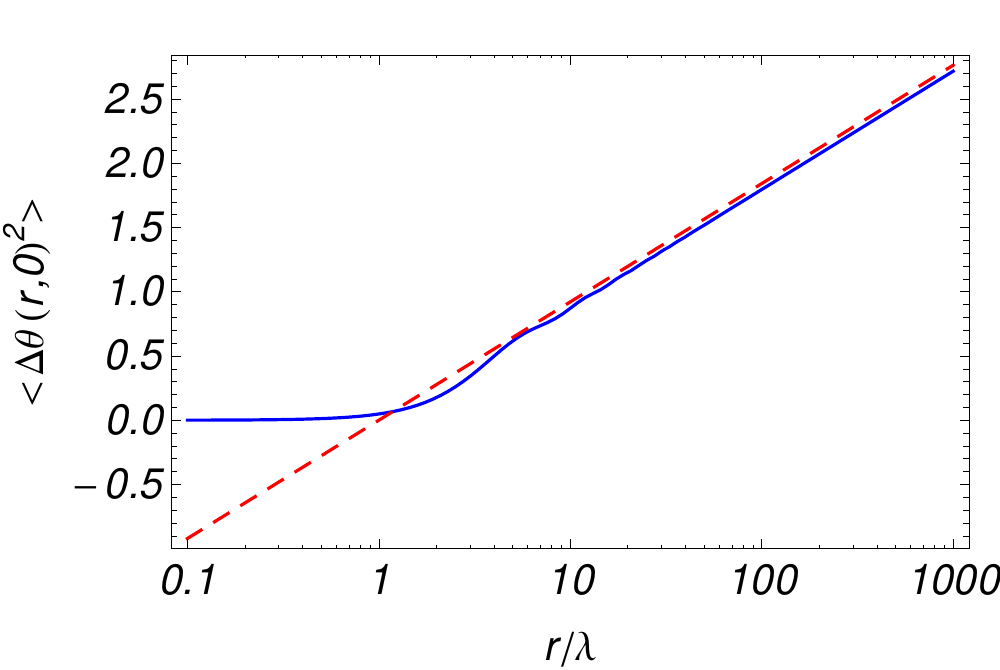}
\caption{Phase fluctuations $\langle\Delta\theta(\textbf{r},0)^2\rangle$ for the homogeneous system. The solid blue curve shows the actual expression given by the integral in Eq. (\ref{hom5}) with $\Lambda=\lambda^{-1}$. For $r\to 0$ the function vanishes as required for the correlation function. The dashed red curve shows the approximation of the integral by the logarithm in Eq. (\ref{hom6}). This is a good and useful approximation for $r\gtrsim 4\lambda$, but cannot capture the correct correlations for smaller $r$. Note that the constant offset for large $r$ does not influence the algebraic scaling of $\rho(r,0)$. The curves can, however, be matched by replacing the ultraviolet cutoff of the integral with $\Lambda\simeq 1.1 \lambda^{-1}$.}
\label{FigHom}
\end{figure}

The integral is conveniently performed in polar coordinates, with angular integration $\frac{1}{2\pi}\int_0^{2\pi}\mbox{d}\phi e^{\rmi q r \cos \phi}=J_0(qr)$, where $J_m(x)$ is the Bessel function \cite{AbrStegun}. We then arrive at
\begin{align}
 \label{hom5}  \langle\Delta\theta(\textbf{r},\textbf{r}')^2\rangle_{\rm hom} =\frac{MT}{\pi n_0} \int_0^\Lambda \frac{\mbox{d}q}{q}\Bigl(1-J_0(q|\textbf{r}-\textbf{r}'|)\Bigr). 
\end{align}
Note that the right hand side does, as is required for the proper correlation function, vanish for $\textbf{r}=\textbf{r}'$ due to $J_m(0)=\delta_{m0}$. To obtain the scaling form of this expression observe that $J_0(qr)$ is positive and of order unity for $qr\leq1$ and small and rapidly oscillating for $qr\geq 1$. Consequently we have
\begin{align}
 \label{hom6} \langle\Delta\theta(\textbf{r},\textbf{r}')^2\rangle_{\rm hom}&\approx \frac{MT}{\pi n_0} \int_{|\textbf{r}-\textbf{r}'|^{-1}}^{\lambda^{-1}} \frac{\mbox{d}q}{q}= \frac{MT}{\pi n_0} \log\Bigl(\frac{|\textbf{r}-\textbf{r}'|}{\lambda}\Bigr).
\end{align}
We compare Eq. (\ref{hom5}) and the result for $|\textbf{r}-\textbf{r}'|\gg\lambda$ in Fig. \ref{FigHom}.

Closer to our analysis, but usually not covered in textbooks, is the treatment of the homogeneous limit in a manifestly radially symmetric setting. Indeed, Eqs. (\ref{hom1})-(\ref{hom6}) should rather be understood as obtained in a cube of size $[0,L]^2$ with $L\to \infty$. In contrast, solving Eq. (\ref{sw10}) for $R\to\infty$ yields the energies $\vare_{qm}\equiv\vare_q=q^2$ with eigenfunctions
\begin{align}
 \label{hom7} \theta_{qm}(\textbf{r}) = \frac{1}{\sqrt{2\pi}} J_m(qr) e^{\rmi m \phi}.
\end{align}
(Note that actually we should write $J_{|m|}(qr)$ . But since $J_m(-x)=(-1)^mJ_m(x)$ and we are free to choose the sign of eigenfunctions, we can also work with $J_m(qr)$.) The properties of the Bessel functions $J_m$ result in the orthogonality and completeness relations
\begin{align}
 \label{hom8} &\int \mbox{d}^2r\ \theta_{qm}^*(\textbf{r}) \theta_{q'm'}(\textbf{r}) =\frac{1}{q} \delta_{mm'}\delta(q-q'),\\
 \label{hom9} &\int_0^\infty \mbox{d}q q \sum_m \theta_{qm}^*(\textbf{r})\theta_{qm}(\textbf{r}') = \frac{1}{r}\delta(r-r')\delta(\phi-\phi').
\end{align}
We also used $\int_0^{2\pi}\mbox{d}\phi e^{-\rmi(m-m')\phi}=2\pi \delta_{mm'}$ and $\sum_m e^{\rmi m (\phi-\phi')}=2\pi\delta(\phi-\phi')$. Writing
\begin{align}
\theta(\textbf{r})=\int_0^\infty \mbox{d}qq\sum_m a_{qm} \theta_{qm}(\textbf{r})
\end{align}
with $a^*_{qm}=a_{q,-m}$ we obtain
\begin{align}
 \label{hom10} S_{\rm ph}[\theta] = \frac{n_0}{2MT} \int_0^\infty \mbox{d}qq\sum_{m} q^2 |a_{qm}|^2.
\end{align}
Note that both sets of eigenfunction for $D_\infty$ are connected by means of 
\begin{align}
e^{\rmi\textbf{q}\cdot\textbf{r}} = \sum_{m=-\infty}^\infty \rmi^m J_m(qr) e^{\rmi m \phi}.
\end{align}

The phase fluctuations of the homogeneous system in this representation read
\begin{align}
  \label{hom11} \langle\Delta\theta(\textbf{r},\textbf{r}')^2\rangle_{\rm hom} = \frac{MT}{n_0} \int_0^\Lambda\mbox{d}qq\sum_m \frac{|\theta_{qm}(\textbf{r})-\theta_{qm}(\textbf{r}')|^2}{q^2}.
\end{align}
Let us first consider $\textbf{r}'=0$. We have
\begin{align}
  \label{hom12} \langle\Delta\theta(\textbf{r},0)^2\rangle_{\rm hom} = \frac{MT}{2\pi n_0} \int_0^\Lambda\frac{\mbox{d}q}{q}\sum_m \Bigl(J_m(qr)-J_m(0)\Bigr)^2.
\end{align}
Furthermore, due to $J_m(0)=\delta_{m0}$ and $ \sum_{m=-\infty}^\infty J_m^2(qr) = 1$ \cite{AbrStegun} we have
\begin{align}
 \label{hom13}  \sum_m[J_m(qr)-J_m(0)]^2 &=2 \Bigl(1-J_0(qr)\Bigr).
\end{align}
Thus we again arrive at
\begin{align}
 \label{hom14} \langle\Delta\theta(\textbf{r},0)^2\rangle_{\rm hom} =\frac{MT}{\pi n_0} \int_0^\Lambda \frac{\mbox{d}q}{q}\Bigl(1-J_0(qr)\Bigr).
\end{align}
The case of $\textbf{r}'\neq0$ in Eq. (\ref{hom5}) can be recovered by applying Graf's addition theorem \cite{AbrStegun}
\begin{align}
 \label{hom15} \sum_{m=-\infty}^\infty J_m(qr)J_m(qr')\cos(m\Delta \phi) = J_0(q|\textbf{r}-\textbf{r}'|).
\end{align}

\section{Trap-averaging}\label{AppTrap}
Here we present the details of the trap-averaging procedure to compute
\begin{align}
 \label{trap1} \bar{g}_1(r) = \int \mbox{d}^2x\ \rho(\textbf{x},\textbf{r}+\textbf{x}).
\end{align}
For notational convenience we set
\begin{align}
 R=1
\end{align}
in this section. We first compute $\bar{g}_1(r)$ by assuming that the spin wave description applies to the whole cloud. Then we estimate the influence of the normal component on the trap averaging. We further present the details of the analysis of the QMC data from Ref. \cite{PhysRevLett.115.010401}.

\subsection{Spin wave contribution}
We define $G_\theta(r,r',\Delta\phi)=e^{-\frac{1}{2}\langle\Delta\theta(\textbf{r},\textbf{r})^2\rangle}$ and arrive at
\begin{align}
   \label{trap2} \bar{g}_1(r) &= n_0 \int_{\mathcal{D}} \mbox{d}^2 x \sqrt{(1-x^2)(1-x'^2)}G_\theta(x,x',\Delta\phi),
\end{align}
where $\textbf{x}'=\textbf{r}+\textbf{x}$, $x=|\textbf{x}|$, $x'=|\textbf{x}'|$, and the integration domain $\mathcal{D}$ is such that the expression under the square root is positive. Furthermore, $\cos \Delta\phi=\textbf{x}\cdot\textbf{x}'/(xx')$, such that $\Delta\phi$ is the angle between $\textbf{x}$ and $\textbf{x}'$. We introduce the angle $\alpha$ through  $\cos\alpha=\textbf{x}\cdot\textbf{r}/(xr)$, i.e., $x'^2=x^2+r^2+2xr\cos\alpha$. Of course, we have
\begin{align}
 \label{trap3} \cos \Delta\phi = \frac{x^2+xr \cos \alpha}{x(x^2+r^2+2xr\cos\alpha)^{1/2}}.
\end{align}
The trap-averaged correlation function becomes
\begin{align}
 \nonumber \bar{g}_{1}(r) &= n_0 \int_0^1 \mbox{d}x x\ \int_{0}^{2\pi}\mbox{d}\alpha\ \theta(1-x'^2)\\
 \label{trap6} &\times \sqrt{(1-x^2)(1-x'^2)} G_{\theta}(x,x',\Delta\phi).
\end{align}
We normalize the function such that $\bar{g}_1(\lambda)=1$ and fit $\bar{g}_1(r)=(r/\lambda)^{-\eta_{\rm trap}}$ in the interval $r\in [\lambda,0.1R]$. Hence we have only one fitting parameter.

\begin{figure}[t]
\centering
\includegraphics[width=8cm]{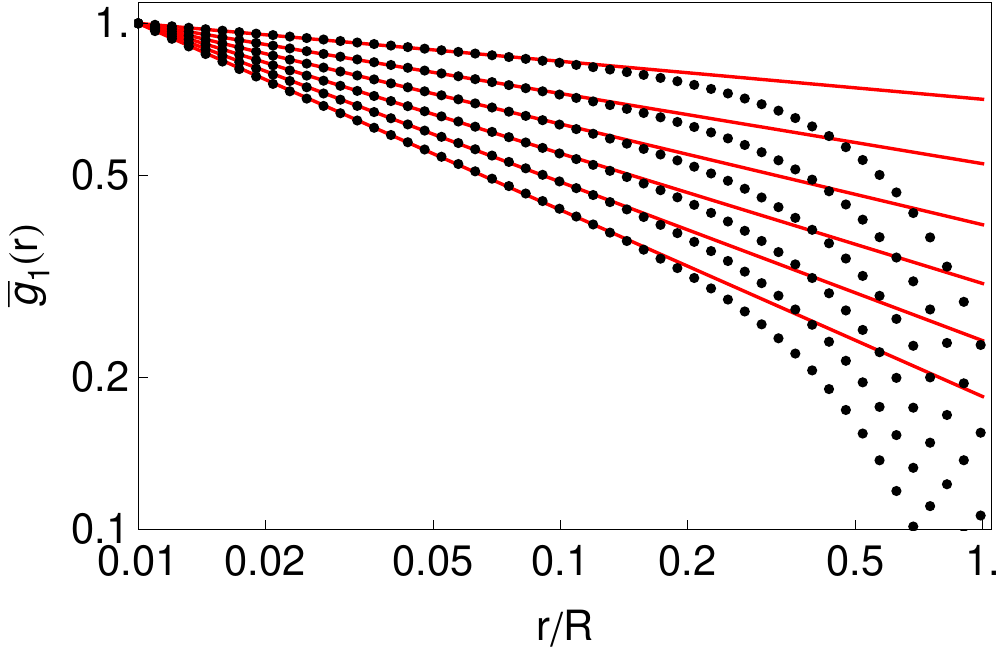}
\caption{Trap-averaged correlation function $\bar{g}_1(r)$ from applying the spin wave description to the whole cloud for $R/\lambda=100$. The curves correspond to $\eta_0=0.04,\ 0.08,\ 0.12,\ 0.16,\ 0.20,\ 0.24$. From the algebraic fit applied to the region $r\in[\lambda,0.1R]$ we obtain $\eta_{\rm trap}=0.075,\ 0.14,\ 0.20,\ 0.27,\ 0.31,\ 0.37$. The dependence of $\eta_{\rm trap}$ on $\eta_0$ and $R/\lambda$ is shown in Fig. \ref{FigSOMeta}.}
\label{FigSOMgbar1sw}
\end{figure}

Within the local correlation approximation we have
\begin{align}
 \label{trap5} G_{\theta,\rm LCA}(x,x',\Delta\phi) = \lambda^{\frac{\eta_0}{2}(\frac{1}{1-x^2}+\frac{1}{1-x'^2})}r^{-\eta_0/\sqrt{(1-x^2)(1-x'^2)}},
\end{align}
which is independent of $\Delta\phi$. From the full spin wave correlation function $\langle \Delta\theta(\textbf{r},\textbf{r}')^2\rangle$ we find
\begin{align}
 \nonumber &G_\theta(x,x',\Delta\phi) = \Bigl(\frac{r^2}{\bar{x}^2_>}\Bigr)^{-\frac{\eta_0}{2\sqrt{(1-x^2)(1-x'^2)}}}\Bigl(\frac{\bar{x}^2_>(1-x^2_<)}{\bar{x}^2_<(1-x^2_>)}\Bigr)^{-\frac{\eta_0}{4}}\\
 \nonumber &\hspace{15mm}\times\Bigl(\frac{\bar{x}^2}{\lambda^2(1-x^2)}\Bigr)^{-\frac{\eta_0}{4(1-x^2)}}\Bigl(\frac{\bar{x}'^2}{\lambda^2(1-x'^2)}\Bigr)^{-\frac{\eta_0}{4(1-x'^2)}}\\
 \label{trap13} &\hspace{15mm}\times e^{-\frac{\eta_0}{2}[\mathcal{R}(x^2,x^2,0)+\mathcal{R}(x'^2,x'^2,0)-2\mathcal{R}(x^2,x'^2,\Delta\phi)]}.
\end{align}
We show $\bar{g}_1(r)$ obtained from Eq. (\ref{trap13}) for a few $\eta_0$ in Fig. \ref{FigSOMgbar1sw} for $R/\lambda=100$.

\begin{figure}[t!]
\centering
\begin{minipage}{0.46\textwidth}
\includegraphics[width=7cm]{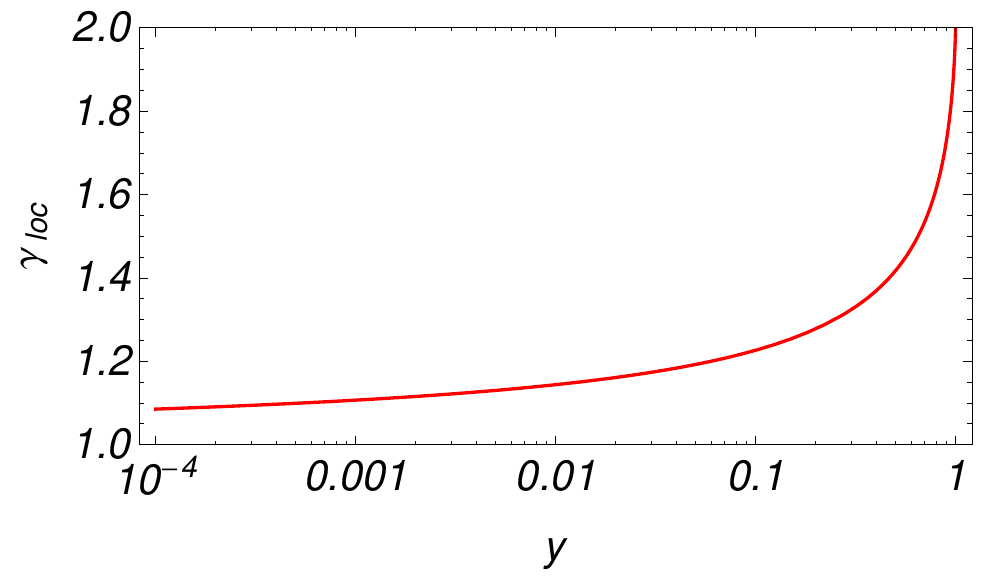}
\includegraphics[width=7cm]{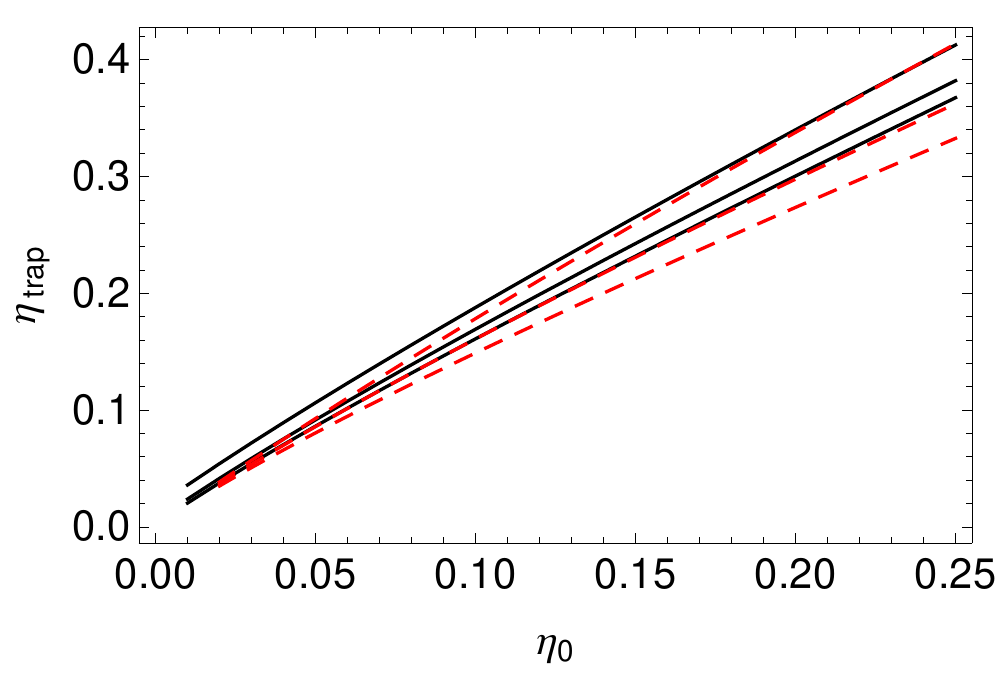}
\caption{\emph{Upper panel.} Local power-law exponent $\gamma_{\rm loc}$ of $L(y)$ from Eqs. (\ref{trap8}) and (\ref{trap9}). The function $\gamma_{\rm loc}(y)$ is bounded from above by two and very slowly approaches unity for $y\to 0$. \emph{Lower panel.} Trap-averaged exponent from $\bar{g}_1(r)$ (solid black curves) for $R/\lambda=20,\ 100,\ 1000$ (from top to bottom), and estimated exponent $\gamma_{\rm loc}(y_{\rm fit})\eta_0$ according to Eq. (\ref{trap11}) with $y_{\rm fit}=(r_{\rm fit}/R)^{-\eta_0}(R/\lambda)^{-\eta_0}$ and $r_{\rm fit}/R=0.1$ (dashed red curves). Given the many approximations that went into computing the estimated exponent, the agreement is satisfactory and thereby explains why $\eta_{\rm trap}>\eta_0$ in general. Further the weak $R/\lambda$-dependence of $\eta_{\rm trap}$ is explained by the smallness of $\eta_0$ such that $y_{\rm fit}$ is almost independent of $R/\lambda$.}
\label{FigSOMeta}
\end{minipage}
\end{figure}

The dependence of $\eta_{\rm trap}$ on $\eta_0$ and $R/\lambda$ can be estimated as follows: For $r\ll R$ (but $r\gg \lambda$) we have $x'\approx x$ and $\Delta\phi\approx 0$. Employing
\begin{align}
 \label{trap5b} G_{\theta,\rm LCA}(x,x,0) &= \Bigl(\frac{r}{\lambda}\Bigr)^{-\eta_0/(1-x^2)}
\end{align}
to leading order in $r\to 0$, we find the trap-averaged correlation function within LCA to be
\begin{align}
 \nonumber \bar{g}_{1,\rm LCA}(r\ll R) &\approx  2\pi n_0 \int_0^1 \mbox{d}x x\  (1-x^2)G_{\theta,\rm LCA}(x,x,0)\\
 &\nonumber =2\pi n_0 \int_0^1 \mbox{d}x x\  (1-x^2) \Bigl(\frac{r}{\lambda}\Bigr)^{-\eta_0/(1-x^2)}\\
 \label{trap7} &= \pi n_0 L\Bigl((r/\lambda)^{-\eta_0}\Bigr),
\end{align}
where we define
\begin{align}
 \nonumber L(y) &= \int_0^1 \mbox{d}t\ (1-t) y^{1/(1-t)} \\
 \label{trap8} &= \frac{1}{2}\Bigl(y + y \log y -(\log y)^2\text{li}(y)\Bigr)
\end{align}
for $0<y<1$. Herein, $\text{li}(y)=\int_0^y \mbox{d}t/\log t$ is the logarithmic integral. Note that $(r/\lambda)^{-\eta_0}$ is slightly below unity due to the small value of $\eta_0$ and $r>\lambda$. We define the local power-law exponent of $L(y)$ by
\begin{align}
 \label{trap9} \gamma_{\rm loc}(y) = \frac{\mbox{d}\log L(y)}{\mbox{d}\log y} = \frac{y}{L(y)}L'(y).
\end{align}
It interpolates between $\gamma_{\rm loc}(0)=1$ and $\gamma_{\rm loc}(1)=2$, see Fig. \ref{FigSOMeta}. We denote the typical fitting range by $r_{\rm fit}$. We then have
\begin{align}
 \label{trap10} \bar{g}_{1,\rm LCA}(r) \simeq \pi n_0 \Bigl(\frac{r}{\lambda}\Bigr)^{-\eta_0\gamma_{\rm loc}(y_{\rm fit})}
\end{align}
with $y_{\rm fit}=(r_{\rm fit}/\lambda)^{-\eta_0} = (r_{\rm fit}/R)^{-\eta_0} (R/\lambda)^{-\eta_0}$. Accordingly,
\begin{align}
 \label{trap11} \eta_{\rm trap}^{(\rm LCA)} \approx \eta_0 \gamma_{\rm loc}(y_{\rm fit}) \in [1,2]\eta_0.
\end{align}
Note that due to the many approximations that went into this equation, a precise match of results cannot be expected. Still, we find satisfying qualitative agreement of the approximative formula (\ref{trap11}) and the results from fitting $\bar{g}_1(r)$, as is displayed in Fig. \ref{FigSOMeta}.

For the actual spin wave correlation function we have
\begin{align}
 \label{trap5c} G_\theta(x,x,0) &= \Bigl(\frac{r}{\lambda}\Bigr)^{-\eta_0/(1-x^2)} (1-x^2)^{\frac{\eta_0}{2(1-x^2)}}
\end{align}
to leading order in $r\to 0$ and deduce
\begin{align}
 \nonumber \bar{g}_{1}(r\ll R) &\approx 2\pi n_0 \int_0^1 \mbox{d}x x\ (1-x^2) G_\theta(x,x,0)\\
 \nonumber &= 2\pi n_0 \int_0^1 \mbox{d}x x\ (1-x^2)^{1+\frac{\eta_0}{2(1-x^2)}} \Bigl(\frac{r}{\lambda}\Bigr)^{-\eta_0/(1-x^2)}\\
 \label{trap15} &= \pi n_0 K\Bigl(\eta_0,(r/\lambda)^{-\eta_0}\Bigr)
\end{align}
with
\begin{align}
 \label{trap16} K(\eta_0,y) = \int_0^1 \mbox{d}t  (1-t)^{1+\frac{\eta_0}{2(1-t)}} y^{1/(1-t)}.
\end{align}
Note that $K(0,y)=L(y)$. Since $\eta_0$ is small, the trap-exponent extracted within LCA is close to the trap-exponent from the actual formula for the phase fluctuations, and thus
\begin{align}
 \eta_{\rm trap} \approx \eta_{\rm trap}^{(\rm LCA)}\approx \eta_0 \gamma_{\rm loc}(y_{\rm fit}) \in [1,2]\eta_0
\end{align}
as well.

We may use Eq. (\ref{trap11}) to estimate the fitting procedure that would yield $\eta_{\rm trap} = 1.1 \eta_0$, i.e., a $10\%$ deviation. For $\gamma_{\rm loc}=1.1$ we need $y_{\rm fit}=0.0005$ or  $y_{\rm fit}=\epsilon^{1/4}$ with $\epsilon=10^{-13}$. For $\eta_0<1/4$ we then have
\begin{align}
 \label{trap17}  \frac{\lambda}{R} \times  \frac{R}{r_{\rm fit}} = \frac{\lambda}{r_{\rm fit}} = y_{\rm fit}^{1/\eta_0} = \epsilon^{1/(4\eta_0)}<\epsilon.
\end{align}
Since $r_{\rm fit}/R\leq 1$ we arrive at
\begin{align}
 \label{trap18} \frac{R}{\lambda} > \frac{1}{\epsilon} = 10^{13}.
\end{align}
For $\lambda=1\mu\text{m}$ we find $R=10000\ \text{km}$, which is of the order of the diameter of the earth.

\subsection{Normal component contribution}
So far we assumed the whole atomic cloud to be captured by the spin wave description. However, the latter only applies to the central superfluid region of the gas. Within the LDA we can estimate the core radius $r_{\rm s}$ from the local superfluid density $n_{\rm s}(r)$ satisfying $n_{\rm s}(r_{\rm s})\lambda^2=4$ and thus find
\begin{align}
 \label{trap19} r_{\rm s}\approx(1-4\eta_0)^{1/2}.
\end{align}
Typical values of $r_{\rm s}$ are a few tens of percent of $R=1$. We may then replace Eq. (\ref{trap2}) according to $\bar{g}_1(r)\to\tilde{g}_1(r)$ with
\begin{align}
   \label{trap20} \tilde{g}_1(r) &= n_0 \int_{\mathcal{D}} \mbox{d}^2 x \sqrt{(1-x^2)(1-x'^2)}\tilde{G}_\theta(x,x',\Delta\phi),
\end{align}
where
\begin{align}
 \label{trap21} \tilde{G}_{\theta}(x,x',\Delta\phi) = \begin{cases} G_\theta(x,x',\Delta\phi) & x<r_{\rm s},\ x'<r_{\rm s}\\ e^{-r/\lambda} & \text{else}\end{cases}.
\end{align}
This constitutes a rather rough first-order estimate. In particular, we approximate the superfluid region as a square with side lengths $2r_{\rm s}$ for computational simplicity. Still it serves to obtain a qualitative picture. Note also that equating the correlation length with $\lambda$ constitutes an approximation that may influence the result. An improved estimate is given by the correlation length $\xi=e^{n\lambda^2/2}/\sqrt{4\pi}$ \cite{Dalibard2011}. However, for $n\lambda^2\gtrsim 1$ this is of order unity, so we will work with $\xi=\lambda$

The translation invariant ansatz $\langle e^{\rmi(\theta(\textbf{r})-\theta(\textbf{r}'))}\rangle\to e^{-|\textbf{r}-\textbf{r}'|/\lambda}$ in the normal component yields an integrand $G_\theta(x,x',\Delta\phi)\sim e^{-r/\lambda}$ which only depends on $r$, but is independent of $x,x',\Delta\phi$. In the typical fitting range we have $r/R\approx 0.1$, so that the contribution of the normal component to $\tilde{g}_1(r)$ is exponentially damped by a prefactor $\approx e^{-0.1R/\lambda}$. This explains that the contribution of the normal component is only visible for relatively small $R/\lambda=\mathcal{O}(10)$.

\begin{figure}[t!]
\centering
\begin{minipage}{0.46\textwidth}
\includegraphics[width=7cm]{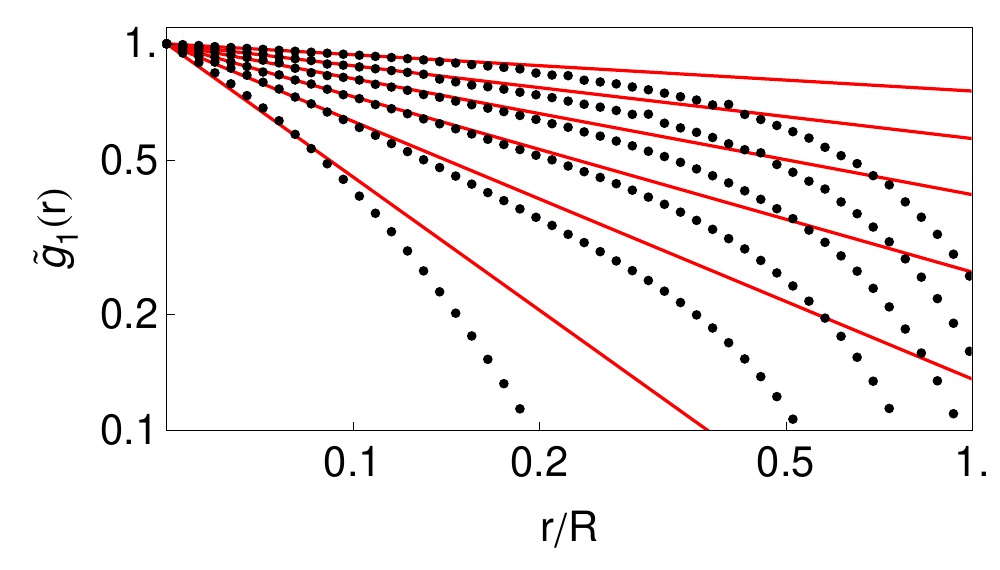}
\includegraphics[width=7cm]{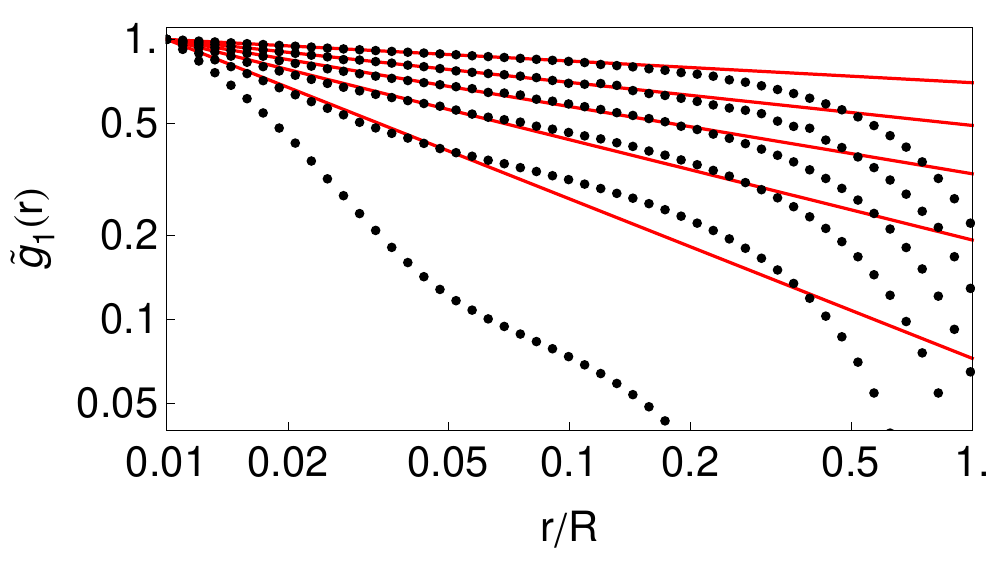}
\includegraphics[width=7cm]{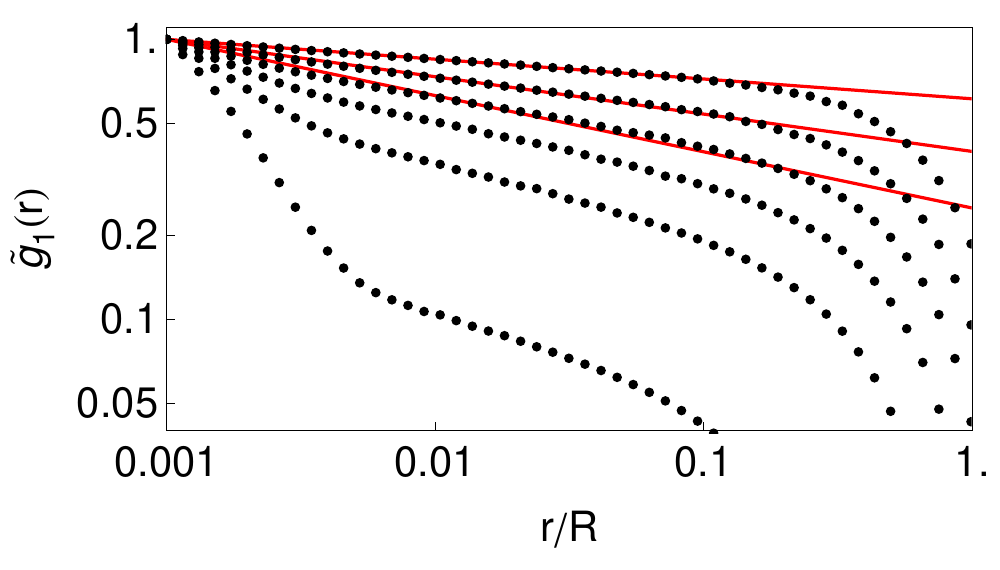}
\caption{Trap-averaged correlation function $\tilde{g}_1(r)$ with the normal component treated according to Eqs. (\ref{trap20}) and (\ref{trap21}). The upper, middle, and lower panel correspond to $R/\lambda=20,\ 100,\ 1000$, i.e., $\hat{\lambda}=0.05,\ 0.01,\ 0.001$. In all panels, the exponents of the central correlation function are chosen as $\eta_0=0.04,\ 0.08,\ 0.12,\ 0.16,\ 0.20,\ 0.24$ (from top to bottom), which is the same as in Fig. \ref{FigSOMgbar1sw}. We fit algebraic decay according to $\tilde{g}_1(r) = (r/\lambda)^{-\tilde{\eta}_{\rm trap}}$ in the interval $r\in[\lambda,0.1R]$ (straight red lines). The obtained trap-exponents are shown by the dashed lines in Fig. 2 in the main text. As $R/\lambda$ and $\eta_0$ increase, the correlation function develops a bimodal structure that cannot be described by a single power-law exponent for $r\leq 0.1R$.}
\label{FigSOMgbar1}
\end{minipage}
\end{figure}

We show the correlation functions $\tilde{g}_1(r)$ for $R/\lambda=20,\ 100,\ 1000$ in Fig. \ref{FigSOMgbar1}. We find that the behavior of $\tilde{g}_1(r)$ for small $r$ and small $\eta_0$ is still well-described by an algebraic decay with exponent $\tilde{\eta}_{\rm trap}$. The latter is above the one fitted from $\bar{g}_1(r)$, thus $\eta_{\rm trap}$ gives a lower bound on the actual exponent that is measured in experiment. The value of $\tilde{\eta}_{\rm trap}$ strongly depends on $R/\lambda$: For $R/\lambda=20,100$, algebraic decay with an increased exponent $\tilde{\eta}_{\rm trap}>\eta_{\rm trap}$ is visible. For $R/\lambda=1000$ we have $\eta_{\rm trap}\approx \tilde{\eta}_{\rm trap}$ for small $\eta_0\lesssim0.15$, whereas $\tilde{g}_1(r)$ is not algebraic for larger $\eta_0$.

\subsection{Quantum Monte Carlo}
The QMC results of Ref. \cite{PhysRevLett.115.010401} for $g_1(r)$ and $\bar{g}_1(r)$ describe a strongly anisotropic 3D trap in accordance with the parameters of the experiment. Here we use this data to compare to the purely 2D problem of the present work. As a consequence, finite size and residual condensation effects from the quasi-2D setup hinder a straightforward quantitative comparison. Still, many qualitative insights can be gained from the comparison.

\begin{figure}[t]
\centering
\includegraphics[width=8cm]{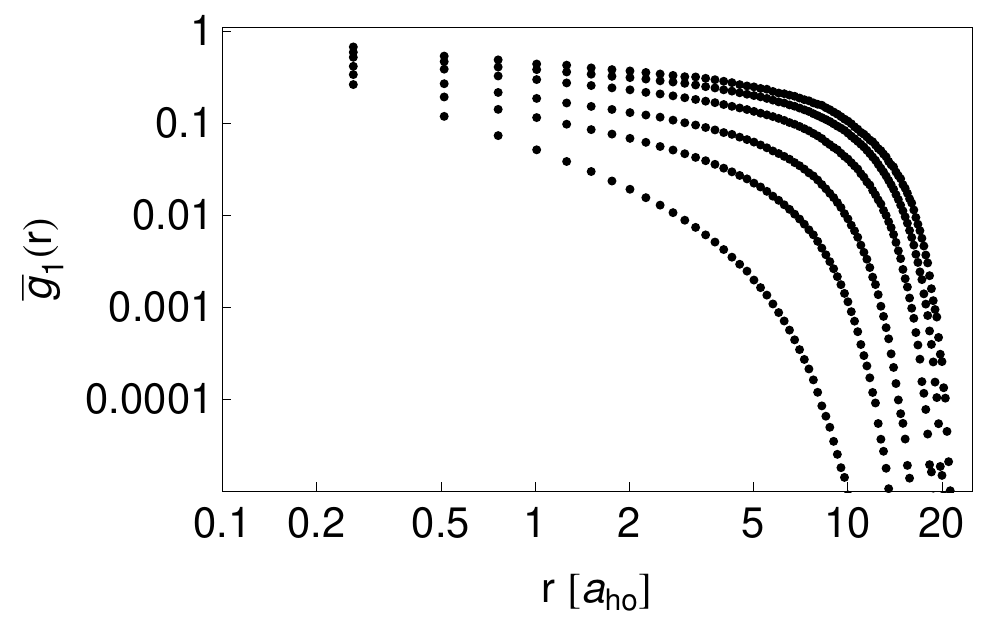}
\caption{Trap-averaged correlation function $\bar{g}_1(r)$ from the QMC calculations of Ref. \cite{PhysRevLett.115.010401} for $\tilde{g}=0.60$. Here $a_{\rm ho}=\sqrt{\hbar/M\omega}$ is the oscillator length in the plane. The data sets correspond to the values displayed in Tab. \ref{Tabelle}. We clearly observe algebraic decay for small $r$ with an increased exponent $\eta_{\rm trap}$. From analogous data for the central correlation function $g_1(r)$ we extract the scaling exponent $\eta_0$ of the QMC results. This yields the data points shown in Fig 2 and Tab. \ref{Tabelle}.}
\label{FigSOMgbar1qmc}
\end{figure}

In Fig. \ref{FigSOMgbar1qmc} we show the trap-averaged correlation function $\bar{g}_1(r)$ obtained from QMC. We observe an algebraic decay with increased exponent $\eta_{\rm trap}$. In order to determine $\eta_0$, two methods qualify: (i) One may use the central density $n_0$ of the gas and then estimate the central superfluid density $n_{\rm s0}$ from the results of classical 2D Monte Carlo simulation such as given in Ref. \cite{PhysRevA.66.043608}. This, in turn, allows to compute $\eta=MT/(2\pi n_{\rm s0})$. As the central density in the present quasi-2D QMC simulations differs from the classical 2D result, see Fig. 2 of Ref. \cite{PhysRevLett.116.045303}, the residual influence of transverse excitations needs to be taken into account, which can be modelled with mean field theory \cite{2008EL.....8230001H,Holzmann2010}. (ii) One may also determine $\eta_0$ from the algebraic scaling of $g_1(r)$. The values of $n_{\rm s0}\lambda_T^2$, $\eta=1/(n_{\rm s0}\lambda_T^2)$, $\eta_0$ and $\eta_{\rm trap}$, together with the estimate of the fitting error, are displayed in Tab. \ref{Tabelle}.

\begin{table}[t]
\begin{tabular}{|c||c|c||c|c|}
\hline\ $t=T/T_{\rm BEC}^0$\ &\ $n_{\rm s0}\lambda_T^2$\ &\ $\eta=1/(n_{\rm s0}\lambda_T^2)$ \ & $\eta_0$ & $\eta_{\rm trap}$ \\ 
\hline\hline\ 0.45\ &\ 14\ &\ 0.07\ & \ 0.10(3)\ &\ 0.28(5)\ \\ 
\hline\ 0.50\ &\ 12.6\ &\ 0.079\ & \ 0.11(3)\ &\ 0.29(5)\ \\ 
\hline\ 0.56\ &\ 10\ &\ 0.10\ & \ 0.15(3)\ &\ 0.43(5)\ \\ 
\hline\ 0.63\ &\ 8.3\ &\ 0.12\ & \ 0.18(3)\ &\ 0.59(5)\ \\ 
\hline\ 0.67\ &\ 6.5\ &\ 0.15\ & \ 0.22(3)\ &\ 0.90(10)\ \\ 
\hline\ 0.71\ &\ 4.2\ &\ 0.23\ & \ 0.27(3)\ &\ 1.38(10)\ \\ 
\hline 
\end{tabular} 
\caption{QMC results of Ref. \cite{PhysRevLett.115.010401}.  The first column gives the temperature in units of the condensation temperature of an ideal gas with the same particle number, and is shown here in order to simplify comparison with the reference. The scaling exponents $\eta_0$ and $\eta_{\rm trap}$ are determined from the algebraic decay of $g_1(r)$ and $\bar{g}_1(r)$, respectively. For $\eta_{\rm trap}$ we use the values given in Ref. \cite{PhysRevLett.115.010401}. The errors of $\eta_0$ and $\eta_{\rm trap}$ estimate the fitting error which mostly results from the fitting range dependence, which we choose in the range $(2\dots 5)a_{\rm ho}$. We also show the estimates for the central superfluid density $n_{\rm s0}$ (multiplied by the thermal wavelength $\lambda_T^2 = 2\pi \hbar^2/(M k_{\rm B}T)$) and the corresponding exponent $\eta=1/(n_{\rm s0}\lambda_T^2)$.}
\label{Tabelle}
\end{table}

We observe that the results for $\eta=1/(n_{\rm s0}\lambda_T^2)$ and $\eta_0$ determined through methods (i) and (ii) are not fully consistent. This is to be expected due to the strong influence of finite size effects leading to deviations from the relation $\eta=\eta_0$ which is valid for the homogeneous system in the thermodynamic limit. Even if the central density $n_0$ of the trapped system and the superfluid fraction $n_{\rm s}/n$ of the homogeneous system are known, the estimate for $n_{\rm s0}$ need not coincide with the central superfluid density of the trapped system, because the latter as a nonlocal observable receives larger finite size corrections and thus need not be converged locally to the value expected from the homogeneous case and LDA. For our analysis we thus employ the value of $\eta_0$ from (ii). It should give a good estimate on how the scaling of $g_1(r)$ translates into the scaling of $\bar{g}_1(r)$, because both observables are subject to similar finite size corrections.

\section{Some sums of Legendre and Jacobi polynomials}\label{AppTools}

The summation techniques presented here are inspired by the works \cite{EnglisPeetre,Gustavsson} from the mathematical literature.

\subsection{Legendre and Jacobi polynomials}\label{AppLegendre}
In this section we collect properties of the Jacobi polynomials $P_n^{(\alpha,\beta)}$ for $\alpha=|m|$ and $\beta=0$. For the polynomials to be normalizable we require $\alpha>-1$. In order to simplify notation we write $P_n^{(m,0)}$ and implicitly assume $m\geq 0$ in this section.

Before discussing the more general case of Jacobi polynomials with $m\in\mathbb{Z}$, we first consider the case of $m=0$. The $n$th Legendre polynomial $P_n(x)=P_n^{(0,0)}(x)$ is a regular solution of the Legendre differential equation
\begin{align}
 \label{leg1} -(1-x^2)\frac{\mbox{d}^2}{\mbox{d}x^2}P_n(x)+2x\frac{\mbox{d}}{\mbox{d}x}P_n(x)=n(n+1)P_n(x),
\end{align}
with $n=1,2,\dots$ an integer. For $n=0$ we set $P_0(x)=1$ so that Eq. (\ref{leg1}) is still satisfied. The $P_n$ are orthogonal according to
\begin{align}
 \label{leg2} \int_{-1}^1 \mbox{d}x\ P_n(x) P_{n'}(x) = \frac{2}{2n+1}\delta_{nn'}.
\end{align}
They further satisfy the completeness relation
\begin{align}
 \label{leg3} \sum_{n=0}^\infty \frac{2n+1}{2} P_n(x) P_n(x') = \delta(x-x').
\end{align}  
For our purposes it is more convenient to work with the variable $s=(1-x)/2$. Equation (\ref{leg1}) can then be written as
\begin{align}
 \label{leg4} -\frac{\mbox{d}}{\mbox{d}s}\Bigl[s(1-s)\frac{\mbox{d}}{\mbox{d}s} P_n(1-2s)\Bigr]=n(n+1)P_n(1-2s).
\end{align}
In terms of the operator (\ref{sw12}) this equation reads $\hat{D}^{(0)}P_n(1-2s)=4n(n+1)P_n(1-2s)$, which proofs formula (\ref{sw13}) for $m=0$. The orthogonality and completeness relations in terms of $s$ read
\begin{align}
  \label{leg5} \int_0^1 \mbox{d}s\ P_n(1-2s)P_{n'}(1-2s) &= \frac{1}{2n+1}\delta_{nn'}
\end{align}
and
\begin{align}
  \label{leg6} \sum_{n=0}^\infty (2n+1) P_n(1-2s)P_n(1-2s') &= \delta(s-s').
\end{align}
For $s=0$ we have $P_n(1)=1$ for all $n$.

The Jacobi polynomials $P_n^{(m,0)}$ are regular solutions of the Jacobi differential equation
\begin{align}
 \nonumber &-(1-x^2)\frac{\mbox{d}^2}{\mbox{d}x^2}P_n^{(m,0)}(x) +\Bigl(m+(m+2)x\Bigr)\frac{\mbox{d}}{\mbox{d}x}P_n^{(m,0)}(x) \\
 \label{jac1} &= n(n+m+1)P_n^{(m,0)}(x).
\end{align}
with $n=1,2,\dots$ an integer. In terms of the variable $s=(1-x)/2$ we have
\begin{align}
 \nonumber &-\frac{\mbox{d}}{\mbox{d}s}\Bigl[s(1-s)\frac{\mbox{d}}{\mbox{d}s}P_n^{(m,0)}(1-2s)\Bigr]\\
 \nonumber &-m(1-s)\frac{\mbox{d}}{\mbox{d}s}P_n^{(m,0)}(1-2s)\\
 \label{jac2} &=n(n+m+1)P_n^{(m,0)}(1-2s).
\end{align}
The polynomials satisfy the orthogonality and completeness relations
\begin{align}
 \label{jac3} \int_0^1 \mbox{d}s\ s^m P_n^{(m,0)}(1-2s)P_{n'}^{(m,0)}(1-2s) = \frac{\delta_{nn'}}{2n+m+1}
\end{align}
and
\begin{align}
 \nonumber &\sum_{n=0}^\infty (2n+m+1)(ss')^{m/2} P_n^{(m,0)}(1-2s)P_n^{(m,0)}(1-2s')\\
  \label{jac4}&=\delta(s-s').
\end{align}

The orthogonality relation (\ref{jac3}) suggests to consider the functions
\begin{align}
 \label{jac6} Q_{nm}(s) = s^{m/2} P_n^{(m,0)}(1-2s).
\end{align}
Due to Eq. (\ref{jac2}) we then arrive at
\begin{align}
 \nonumber &-\frac{\mbox{d}}{\mbox{d}s}\Bigl[s(1-s)\frac{\mbox{d}}{\mbox{d}s}Q_{nm}(s)\Bigr]+\frac{m^2}{4}\frac{1-s}{s}Q_{nm}(s) \\
 \label{jac7}  &=\Bigl(\frac{m}{2}+n(n+m+1)\Bigr)Q_{nm}(s).
\end{align}
This is equivalent to $\hat{D}^{(m)}Q_{nm}(s)=[2m+4n(n+m+1)]Q_{nm}(s)$ and thus yields Eq. (\ref{sw13}). From Eqs. (\ref{jac4}) and (\ref{jac5}) we obtain
\begin{align}
 \label{jac8} &\int_0^1 \mbox{d}s\ Q_{nm}(s) Q_{n'm}(s)=\frac{\delta_{nn'}}{2n+m+1},\\
 \label{jac9} & \sum_{n=0}^\infty (2n+m+1)Q_{nm}(s)Q_{nm}(s') = \delta(s-s').
\end{align}

The Jacobi polynomials have the explicit expression
\begin{align}
 \label{jac5} P_n^{(m,0)}(1-2s) = \sum_{k=0}^n (-1)^k \binom{n}{k}\binom{n+m+k}{n} s^k.
\end{align}
For large $n$ they satisfy
\begin{align}
 P_n^{(m,0)}\Bigl(\cos(z/n)\Bigr)\sim P_n^{(m,0)}\Bigl(1-\frac{z^2}{2n^2}\Bigr) \sim \Bigl(\frac{2n}{z}\Bigr)^mJ_m(z),
\end{align}
which implies
\begin{align}
 s^{m/2} P_n^{(m,0)}(1-2s) \sim J_m(2n\sqrt{s})
\end{align}
for large $n$ and small $s$. For $s=0$ we have $P_n^{(m,0)}(1)=\binom{n+m}{n}$.

\subsection{Hypergeometric function and Pochhammer symbol}\label{AppHyper}
We define the hypergeometric function ${}_2F_1$ and discuss a few properties of the function that are relevant for our analysis.

The hypergeometric differential equation reads
\begin{align}
 \label{hyper1} 0 = s(1-s)w''(s)+\Bigl[c-(a+b+1)s\Bigr]w'(s)-abw(s),
\end{align}
where $a,b,c$ are parameters which determine the properties of $w(s)= w(a,b,c;s)$. A solution to this equation for $|s|<1$ is given by $w(s)={}_2F_1(a,b,c,s)$, where
\begin{align}
 \label{hyper2} {}_2F_1 (a,b,c,s) = 1+\sum_{n=1}^\infty \frac{(a)_n(b)_n}{(c)_n}\frac{s^n}{n!}.
\end{align}
Here we define the (rising) Pochhammer symbol $(a)_n$ for $n>0$ by
\begin{align}
 \label{hyper3} (a)_n = a(a+1)\dots(a+n-1) =\prod_{i=0}^{n-1} (a+i).
\end{align}
For convenience we set $(a)_0=1$, so that we may also write ${}_2F_1(a,b,c,s) = \sum_{n=0}^\infty \frac{(a)_n(b)_n}{(c)_n}\frac{s^n}{n!}$. For later reference note that
\begin{align}
 \label{hyper4} (a)_n =\frac{\Gamma(a+n)}{\Gamma(a)},\ \frac{(a+1)_n}{(a)_n} = \frac{a+n}{a},\ (1)_n=n!.
\end{align}
If $a$ is a negative integer, say $a=-k$ with $k=1,2,\dots$, then $(a)_n=0$ for $n\geq k+1$. Consequently ${}_2F_1(a,b,c,s)$ is a polynomial if either $a$ or $b$ are a negative integer, whereas the expression is not defined for $c$ being a negative integer. We also have 
\begin{align}
 \label{hyper4b} \sum_{n=0}^\infty (a)_n\frac{s^n}{n!} = (1-s)^{-a}
\end{align}
for $|s|<1$. Further note that ${}_2F_1(a,b,c,s)$ is symmetric with respect to exchanging $a$ and $b$.

Besides $w(s)={}_2F_1(a,b,c,s)$ there are other solutions to Eq. (\ref{hyper1}). For our purposes a very simplistic approach will be sufficient. We seek two linearly independent solutions of Eq. (\ref{hyper1}). Potential candidates are any two out of the following six functions:
\begin{align}
 \nonumber w_1(s) & = {}_2F_1(a,b,c,s),\\
 \nonumber w_2(s) &= s^{1-c}{}_2F_1(1+a-c,1+b-c,2-c,s),\\
 \nonumber w_3(s) &= {}_2F_1(a,b,1+a+b-c,1-s),\\
 \nonumber w_4(s) &= (1-s)^{c-a-b}{}_2F_1(c-a,c-b,1+c-a-b,1-s),\\
 \nonumber w_5(s) &= s^{-a}{}_2F_1(a,1+a-c,1+a-b,s^{-1}),\\
 \label{hyper5} w_6(s) &= s^{-b} {}_2F_1(b,1+b-c,1+b-a,s^{-1}).
\end{align}
These functions have different regularity properties at $s=0,1,\infty$.

Many relations for the function ${}_2F_1(a,b,c,s)$ exist and are tabulated, for instance, in Ref. \cite{AbrStegun}. Here we mention the Euler transformation
\begin{align}
 \label{hyper6} {}_2F_1(a,b,c,s) = (1-s)^{c-a-b}{}_2F_1(c-a,c-b,c,s).
\end{align}
The formula can be applied when $a,b,c$ are such that the result is not divergent. 
We have the Euler integral representation
\begin{align}
 \label{hyper9} {}_2F_1(a,b,c,s) &= \frac{\Gamma(c)}{\Gamma(b)\Gamma(c-b)} \\
 \nonumber  &\times \int_0^1 \mbox{d}x\ x^{b-1}(1-x)^{c-b-1}(1-sx)^{-a},
\end{align}
from which we easily obtain Gau\ss' formula
\begin{align}
 \label{hyper10} {}_2F_1(a,b,c,1) = \frac{\Gamma(c)\Gamma(c-a-b)}{\Gamma(c-a)\Gamma(c-b)}
\end{align}
for $c>a+b$. The derivative of the hypergeometric function is given by
\begin{align}
  \label{hyper11} \frac{\mbox{d}}{\mbox{d}s} {}_2F_1(a,b,c,s) = \frac{ab}{c} {}_2F_1(a+1,b+1,c+1,s).
\end{align}

\subsection{Fundamental solutions for Sturm--Liouville operators}\label{AppSturm}

In this section we consider the Sturm--Liouville differential operator
\begin{align}
 \label{sturm1} L_s=\frac{\mbox{d}}{\mbox{d}s}\Bigl[p(s)\frac{\mbox{d}}{\mbox{d}s}\Bigr] + q(s),
\end{align}
where $p(s)$ and $q(s)$ are functions of the one-dimensional variable $s$. We want to construct the fundamental solution of this operator, i.e., the function $G(s,t)$ satisfying
\begin{align}
 \label{sturm2} L_s G(s,t) = -\delta(s-t).
\end{align}

To construct $G$ we assume that the equation $L_sf(s)=0$ has two linearly independent solutions $u(s)$ and $v(s)$. Define
\begin{align}
 \label{sturm3} G(s,t) &= \begin{cases} u(t)v(s) &( t\leq s) \\ u(s)v(t) & (s\leq t) \end{cases}\\
 \nonumber &=u(t)v(s)\theta(s-t)+u(s)v(t)\theta(t-s).
\end{align}
We show that this is indeed already the desired function up to a constant prefactor. We have
\begin{align}
 \nonumber L_sG(s,t)  &= \frac{\mbox{d}}{\mbox{d}s} \Bigl[ p(s)u(t)v'(s)\theta(s-t)+p(s)u(t)v(s)\delta(s-t)\\
 \nonumber &+p(s)u'(s)v(t)\theta(t-s)-p(s)u(s)v(t)\delta(t-s)\Bigr] \\
 \label{sturm4}&+ q(s)G(s,t).
\end{align}
We observe the terms multiplying the delta functions to cancel. Eliminating them we are left with
\begin{align}
 \nonumber &L_sG(s,t) \\
 \nonumber &= \theta(s-t)u(t)\frac{\mbox{d}}{\mbox{d}s}\Bigl[p(s)\frac{\mbox{d}}{\mbox{d}s}v(s)\Bigr]+p(s) u(t)v'(s)\delta(s-t)\\
 \nonumber &+\theta(t-s)v(t)\frac{\mbox{d}}{\mbox{d}s}\Bigl[p(s)\frac{\mbox{d}}{\mbox{d}s}u(s)\Bigr]-p(s)u'(s)v(t)\delta(t-s)\\
 \label{sturm5} &+q(s)\Bigl[u(t)v(s)\theta(s-t)+u(s)v(t)\theta(t-s)\Bigr].
\end{align}
In this express, however, since $u(s)$ and $v(s)$ are zero modes of $L_s$, the terms proportional to the $\theta$-functions cancel as well, and we eventually arrive at
\begin{align}
 \label{sturm6} L_sG(s,t) = -\text{pW}(s) \delta(s-t),
\end{align}
where the function $\text{pW}(s)$ is the product of $p(s)$ and the Wronskian of $u$ and $v$, i.e.,
\begin{align}
 \label{sturm7} \text{pW}(s) = \Bigl[ p(s)u'(s)\Bigr]v(s) -u(s) \Bigl[ p(s)v'(s)\Bigr].
\end{align}
We claim that $\text{pW}(s)$ is constant. To see this note that
\begin{align}
 \nonumber \frac{\mbox{d}}{\mbox{d}s}\text{pW}(s) ={}& -\Bigl[q(s)u(s)\Bigr]v(s)+u(s)\Bigl[q(s)v(s)\Bigr]\\
 \label{sturm8} &+p(s)u'(s) v'(s)-u'(s)p(s)v'(s) =0.
\end{align}
Hence $\text{pW}(z)=C$ is a constant. By a proper rescaling of $G(s,t)$ we can assume $C=1$ and thus arrive at
\begin{align}
 \label{sturm9} L_sG(s,t) = -\delta(s-t).
\end{align}

\subsection{Summation formulas from Green functions}\label{AppSums}
In this section we compute sums of the type
\begin{align}
 \label{sum1} \sum_n \frac{2n+m+1}{\frac{m}{2}+n(n+m+1)}P_n^{(m,0)}(1-2s)P_n^{(m,0)}(1-2t)
\end{align}
for fixed $m\geq0$. In regard of Eq. (\ref{cor8}) it is a great benefit that the $n$-sum can be evaluated analytically, as this leaves us with the $m$-sum, which is readily evaluated numerically.

The basic principle that we employ here to compute sums of the type (\ref{sum1}) is based on the fact, that the eigenfunctions $\varphi_n(s)$ of a self-adjoint operator $D$ form a complete set of functions, namely
\begin{align}
 \label{sum2} \sum_n \varphi_n(s)\varphi^*_n(t) = \delta(s-t).
\end{align}
As a consequence, the function $g(s,t)$ defined by
\begin{align}
 \label{sum3} g(s,t) = - \sum_n \frac{\varphi(s)\varphi^*(t)}{\vare_n},
\end{align}
where $\vare_n$ is the eigenvalue of $\varphi_n$ according to $D \varphi_n =\vare_n \varphi_n$, is a fundamental solution of $D$, i.e.
\begin{align}
 \label{sum4} D_s g(s,t) = -\delta(s-t).
\end{align}
On the other hand, if we know the fundamental solution of $D$ by some other means, we immediately obtain the result of the summation on the right hand side of Eq. (\ref{sum3}). 

In our case, the operator $D$ is given by $\hat{D}^{(m)}$ in Eq. (\ref{sw12}). As the latter one is of Sturm--Liouville type we can construct its fundamental solution from the method presented in App. \ref{AppSturm}. Furthermore, the sum involved in Eq. (\ref{sum3}) is (up to a prefactor) precisely that of Eq. (\ref{sum1}). We treat the cases $m=0$ and $m>0$ separately, as they come in somewhat different flavours. It is convenient to rescale the operator $\hat{D}^{(m)}$ by a factor of $4$ according to
\begin{align}
 \label{sum5} \hat{D}^{(m)} \to D^{(m)}= \frac{1}{4} \hat{D}^{(m)} = \frac{\mbox{d}}{\mbox{d}s} \Bigl[p(s)\frac{\mbox{d}}{\mbox{d}s}\Bigr]+q(s)
\end{align}
with 
\begin{align}
p(s)=-s(1-s),\ q(s) = \frac{m^2}{4}\frac{1-s}{s}.
\end{align}

For $m=0$ and $0<s,t<1$ we define
\begin{align}
 \label{sum6} g^{(0)}(s,t) = -\sum_{n=1}^\infty \frac{2n+1}{n(n+1)}P_n(1-2s)P_n(1-2t).
\end{align}
Since $D^{(0)}_sP_n(1-2s)=n(n+1)P_n(1-2s)$ due to Eq. (\ref{leg4}) we have
\begin{align}
 \nonumber D^{(0)}_sg(s,t) &= -\sum_{n=1}^\infty (2n+1)P_n(1-2s)P_n(1-2t) \\
 \label{sum7}  &= 1-\delta(s-t).
\end{align}
Thus it turns out that $g^{(0)}(s,t)$ is not precisely the fundamental solution of $D^{(0)}$ as the sums starts at $n>0$. (Recall that we exclude $(n,m)=(0,0)$ from the summation over $n$ and $m$.) However, once the Green function of $D^{(0)}$ is known, it will be easy to construct $g^{(0)}(s,t)$.

To obtain the fundamental solution of $D^{(0)}$ we seek two linearly independent zero modes of the operator, i.e., functions $f=u_0,v_0$ satisfying
\begin{align}
 \label{sum8} 0 =D^{(0)}_s f(s) = -[s(1-s)f'(s)]'.
\end{align}
Obviously, these are given by
\begin{align}
 \label{sum9} u_0(s)=1,\ v_0(s) = \log\Bigl(\frac{s}{1-s}\Bigr).
\end{align}
We have chosen the prefactors of $u_0$ and $v_0$ such that $\text{pW}(s)=-s(1-s)(u_0'v_0-u_0v_0')=1$ in Eq. (\ref{sturm7}). The fundamental solution of $D^{(0)}$ is then given by
\begin{align}
\label{sum10} G(s,t) = \begin{cases} u_0(t)v_0(s) & (t\leq s)\\ u_0(s)v_0(t) & (s\leq t) \end{cases} = \begin{cases} \log(\frac{s}{1-s}) & (t\leq s)\\ \log(\frac{t}{1-t}) & (s\leq t)\end{cases}.
\end{align}
From Eq. (\ref{sum7}) we conclude that $g^{(0)}(s,t)=G(s,t)+h(s,t)$ where $h(s,t)$ is a function satisfying
\begin{align}
 \label{sum11} D^{(0)}_sh(s,t) = 1.
\end{align}
The solution to this inhomogeneous equation, however, can be constructed with the help of the Green function $G(s,t)$ and a bit of elaborate guessing. Since $h(s,t)$ is symmetric in $s$ and $t$, we may assume $h(s,t)=C+h(s)+h(t)$ with a constant $C$ and
\begin{align}
 \label{sum12} h(s) = -\int_0^1 \mbox{d}x\ G(s,x) = \log(1-s).
\end{align}
We then arrive at
\begin{align}
 \label{sum13} g^{(0)}(s,t) = \begin{cases} C+\log[s(1-t)] & (t\leq s)\\ C+\log[t(1-s)] & (s\leq t)\end{cases}.
\end{align}
It is now easily seen numerically from summing Eq. (\ref{sum6}) for fixed $s,t$ that $C=1$. We summarize the final result as
\begin{align}
 \nonumber F_0^{(\infty)}(s,t) &= \sum_{n=1}^\infty \frac{2n+1}{n(n+1)}P_n(1-2s)P_n(1-2t) \\
 \label{sum14} &=  \begin{cases} -1-\log[s(1-t)] & (t\leq s)\\ -1-\log[t(1-s)] & (s\leq t)\end{cases}.
\end{align}
These sum formulas can also be found in Refs. \cite{EnglisPeetre} and \cite{Gustavsson}, where they are derived in a slightly different manner. These works also confirm that $C=1$, which supplements our somewhat unsatisfactory numerical determination of $C$. Eq. (\ref{sum14}) can also be applied in the limit that either $s$ or $t$ approaches zero.

We proceed with the case $m>0$. For $0<s,t<1$ we define
\begin{align}
 \label{sum15} g^{(m)}(s,t) = - {}& \sum_{n=0}^\infty \frac{2n+m+1}{\frac{m}{2}+n(n+m+1)} Q_{nm}(s)Q_{nm}(t)
\end{align}
with $Q_{nm}(s)=s^{m/2}P_n^{(m,0)}(1-2s)$ from Eq. (\ref{jac6}). Due to $D^{(m)}Q_{nm}=(\frac{m}{2}+n(n+m+1))Q_{nm}$ we have
\begin{align}
 \nonumber D^{(m)}_sg^{(m)}(s,t) &= -\sum_{n=0}^\infty (2n+m+1)Q_{nm}(s)Q_{nm}(t) \\
 \label{sum16}  &= -\delta(s-t).
\end{align}
Thus $g^{(m)}(s,t)$ is the fundamental solution of the operator $D^{(m)}$. Due to the sum starting at $n=0$, the correspondence is less subtle than in the case of $m=0$. We first determine the zero modes of $D^{(m)}$, i.e., functions $f$ satisfying
\begin{align}
 \label{sum17} D^{(m)}f(s) = \Bigl(-\frac{\mbox{d}}{\mbox{d}s}\Bigl[s(1-s)\frac{\mbox{d}}{\mbox{d}s}\Bigr]+\frac{m^2}{4}\frac{1-s}{s}\Bigr)f(s)=0.
\end{align}
Introducing the function $g$ via $f(s)=s^{m/2}g(s)$ we find
\begin{align}
  \nonumber 0 &=\Bigl(\frac{\mbox{d}}{\mbox{d}s}\Bigl[s(1-s)\frac{\mbox{d}}{\mbox{d}s}\Bigr]+m(1-s)\frac{\mbox{d}}{\mbox{d}s}-\frac{m}{2}\Bigr)g(s)\\
 \label{sum18} &=s(1-s)g''(s) +\Bigl[(m+1)-(m+2)s\Bigr]g'(s)-\frac{m}{2}g(s).
\end{align}
This coincides with the hypergeometric differential equation (\ref{hyper1}) for $a+b+1=m+2$, $ab=m/2$, $c=m+1$. The solution of $a_m+b_m=m+1$ and $a_mb_m=m/2$ reads
\begin{align}
 \label{sum19} a_m &=\frac{m+1}{2}+\frac{\sqrt{m^2+1}}{2},\\
 \label{sum20} b_m &=\frac{m+1}{2}-\frac{\sqrt{m^2+1}}{2}.
\end{align}
We investigate the six solutions from Eq. (\ref{hyper5}) to construct the zero modes $g=u_m,\tilde{v}_m$ of the differential operator defined by Eq. (\ref{sum18}). We observe that besides $w_1(s)$, $w_3(s)=w_4(s)$ is well-defined and considerably simple. Thus we choose
\begin{align}
 \label{sum21} u_m(s) &= {}_2F_1(a_m,b_m,m+1,s),\\
 \label{sum22} \tilde{v}_m(s) &={}_2F_1(a_m,b_m,1,1-s).
\end{align}
The zero modes of $D^{(m)}$ are then given by $f_u(s)=s^{m/2}u_m(s)$ and $f_v(s)=s^{m/2}\tilde{v}_m(s)$. The corresponding function $\text{pW}(s)$ from Eq. (\ref{sturm7}) is given by 
\begin{align}
 \label{sum23} \text{pW}(s) &= -s(1-s) \Bigl[f_u'(s)f_v(s)-f_u(s)f_v'(s)\Bigr]\\
 \nonumber &= -s(1-s) s^m\Bigl[ u_m'(s)\tilde{v}_m(s)-u_m(s)\tilde{v}_m'(s)\Bigr].
\end{align}
According to the arguments presented in Sec. \ref{AppSturm}, this expression is a constant. We show below that $\text{pW}(s)\equiv -p_m=-\frac{\Gamma(m+1)}{\Gamma(a_m)\Gamma(b_m)}$. We eventually arrive at the fundamental solution of $D^{(m)}$, and thus Eq. (\ref{sum15}), being given by
\begin{align}
 \label{sum24} g^{(m)}(s,t) = \begin{cases} - \frac{1}{p_m}f_u(t)f_v(s) & (t\leq s) \\ -\frac{1}{p_m} f_u(s) f_v(t) & (s\leq t) \end{cases}.
\end{align}
Dividing by $(st)^{m/2}$ we then arrive at
\begin{align}
 \nonumber &\sum_{n=0}^\infty \frac{2n+m+1}{\frac{m}{2}+n(n+m+1)}P_n^{(m,0)}(1-2s)P_n^{(m,0)}(1-2t)\\
 \label{sum25} &\hspace{10mm}= \begin{cases} \frac{1}{p_m} u_m(t)\tilde{v}_m(s) & (t\leq s)\\ \frac{1}{p_m}u_m(s)\tilde{v}_m(t) & (s\leq t)\end{cases}
\end{align}
for $0<s,t<1$. In the following we bring the right hand side into an analytically more accessible form.

The function $u_m$ is well-behaved at the origin, where it satisfies $u_m(0)=1$. In contrast, $\tilde{v}_m(s)$ diverges like $s^{-m}$ for $s\to 0$. To obtain a function which is regular at the origin we employ Euler's transformation (\ref{hyper6}) to write
\begin{align}
 \label{sum26} \tilde{v}_m(s) = s^{-m} \chi_m(s)
\end{align}
with
\begin{align}
 \label{sum27} \chi_m(s) = {}_2F_1(1-a_m,1-b_m,1,1-s).
\end{align}
The function $\chi_m$ is finite at the origin and, using Gau\ss' relation (\ref{hyper10}), we have
\begin{align}
 \label{sum28} \chi_m(0) = \frac{\Gamma(m)}{\Gamma(a_m)\Gamma(b_m)}.
\end{align}
This in turn allows us to compute the constant $\text{pW}(s)=-p_m$ from Eq. (\ref{sum23}): For the derivatives of $u_m$ and $\chi_m$ we use Eq. (\ref{hyper11}) to compute
\begin{align}
 \label{sum29} u_m'(s)&= \frac{m/2}{m+1}{}_2F_1(a_m+1,b_m+1,m+2,s),\\
 \label{sum30} \chi_m'(s) &= \frac{m}{2}{}_2F_1(2-a_m,2-b_m,2,1-s),
\end{align}
where we employed $(1-a_m)(1-b_m)=-m/2$. We then arrive at
\begin{align}
 \nonumber \text{pW}(s) = -s(1-s)\Bigl[{}&u_m'(s)\chi_m(s)+\frac{m}{s}u_m(s)\chi_m(s)\\
 \label{sum31}&-u_m(s)\chi_m'(s)\Bigr].
\end{align}
In particular, in the limit $s\to 0$ we find
\begin{align}
 \label{sum32} \text{pW}(s) \equiv -p_m = -m \chi_m(0) = -\frac{\Gamma(m+1)}{\Gamma(a_m)\Gamma(b_m)}.
\end{align} 
Furthermore, the function $v_m(s)=\chi_m(s)/\chi_m(0)$ approaches unity for $s\to 0$. We conclude that a more transparent parametrization of the sum in Eq. (\ref{sum25}) is given by
\begin{align}
 \nonumber &\sum_{n=0}^\infty \frac{2n+m+1}{\frac{m}{2}+n(n+m+1)}P_n^{(m,0)}(1-2s)P_n^{(m,0)}(1-2t)\\
 \label{sum33} &\hspace{10mm}= \begin{cases} \frac{1}{m}s^{-m} u_m(t)v_m(s) & (t\leq s)\\ \frac{1}{m}t^{-m}u_m(s)v_m(t) & (s\leq t)\end{cases},
\end{align}
with 
\begin{align}
 \label{sum34} u_m(s) &= {}_2F_1(a_m,b_m,m+1,s),\\
 \label{sum35} v_m(s) &= \frac{\Gamma(a_m)\Gamma(b_m)}{\Gamma(m)} {}_2F_1(1-a_m,1-b_m,1,1-s).
\end{align}
Of course, within this parametrization we have $u_m(0)=v_m(0)=1$.

As a next step we study the large-$m$ behavior of $u_m$ and $v_m$. We claim that
\begin{align}
 \label{sum36} u_m(s),\ v_m(s) \stackrel{m\to \infty}{\longrightarrow} \frac{1}{\sqrt{1-s}}.
\end{align}
As a consequence, the right hand side of Eq. (\ref{sum33}) only decays like $1/m$ for large $m$ and thus leads to a logarithmic divergence in $M$ when performing the sum $m=1,2,\dots,M$ for $M\gg1$. To proof Eq. (\ref{sum36}) note that for large $m$ we have
\begin{align}
 \label{sum37} a_m &= m+\frac{1}{2} +\frac{1}{4m}+\dots,\\
  b_m &= \frac{1}{2}-\frac{1}{4m}+\dots
\end{align}
Furthermore, we have
\begin{align}
 \label{sum38} \frac{\Gamma(m+1/2)}{\Gamma(m)} \stackrel{m\to \infty}{\longrightarrow} \sqrt{m}.
\end{align}
For large $m$ the function $u_m$ satisfies
\begin{align}
 \label{sum39} u_m(s) &\to {}_2F_1\Bigl(m,\frac{1}{2},m,s\Bigr)=\sum_{n=0}^\infty\Bigl(\frac{1}{2}\Bigr)_n \frac{s^n}{n!}=\frac{1}{\sqrt{1-s}}.
\end{align}
To compute $\chi_m(s)$ for $m\to \infty$ we employ the Euler integral representation (\ref{hyper9}) to write
\begin{align}
 \nonumber \chi_m(s) &=\frac{1}{\Gamma(b_m)\Gamma(1-b_m)} \\
 \nonumber &\times \int_0^1 \mbox{d}x\ x^{-b_m}(1-x)^{b_m-1}[1-(1-s)x]^{a_m-1}\\
 \label{sum40}  &\to \frac{1}{\Gamma(\frac{1}{2})^2} \int_0^1 \mbox{d}x \frac{[1-(1-s)x]^{m}}{\sqrt{x(1-x)}}.
\end{align}
We expand this expression around $s=0$ according to
\begin{align}
 \nonumber \chi_m(s) &\to \frac{1}{\Gamma(\frac{1}{2})^2} \sum_{n=0}^m \binom{m}{n}s^n \int_0^1 \frac{(1-x)^m (\frac{x}{1-x})^n}{\sqrt{x(1-x)}}\\
 \nonumber &=\frac{1}{\Gamma(\frac{1}{2})^2} \sum_{n=0}^m \binom{m}{n}s^n \frac{1}{m!} \Gamma\Bigl(m-n+\frac{1}{2}\Bigr)\Gamma\Bigl(n+\frac{1}{2}\Bigr)\\
 \label{sum41} &=\frac{1}{\Gamma(\frac{1}{2})}\sum_{n=0}^m \Bigl(\frac{1}{2}\Bigr)_n \frac{s^n}{n!} \frac{\Gamma(m-n+\frac{1}{2})}{\Gamma(m-n+1)}.
\end{align}
Now note that
\begin{align}
  \label{sum42} \frac{\Gamma(m-n+\frac{1}{2})}{\Gamma(m-n+1)} \to \frac{\Gamma(m+\frac{1}{2})}{\Gamma(m+1)} = \frac{1}{m} \frac{\Gamma(m+\frac{1}{2})}{\Gamma(m)} \to  \frac{1}{\sqrt{m}}
\end{align}
such that
\begin{align}
 \label{sum43} \chi_m(s) \stackrel{m\gg1}{\to} \frac{1}{\Gamma(\frac{1}{2})} \frac{1}{\sqrt{m}}\sum_{n=0}^\infty \Bigl(\frac{1}{2}\Bigr)_n \frac{s^n}{n!} = \frac{1}{\sqrt{m\pi(1-s)}}.
\end{align}
Of course, for $s=0$ we also have
\begin{align}
 \label{sum44}  \chi_m(0) = \frac{\Gamma(m)}{\Gamma(a_m)\Gamma(b_m)} \to \frac{\Gamma(m)}{\Gamma(m+\frac{1}{2})\Gamma(\frac{1}{2})} \to \frac{1}{\sqrt{m\pi}},
\end{align}
and thus $v_m(s) \to (1-s)^{-1/2}$ as claimed in Eq. (\ref{sum36}).

\end{document}